\documentclass[floats,floatfix,showpacs,amssymb,prd,twocolumn,superscriptaddress,nofootinbib,nolongbibliography,reprint,preprintnumbers]{revtex4-1}

\usepackage{amssymb,amsmath,verbatim,mathtools,needspace,enumitem,etoolbox,graphicx,physics,microtype,afterpage,xspace,tabularx,lmodern,multirow,braket}
\usepackage{gensymb}
\usepackage[normalem]{ulem}
\usepackage[dvipsnames, usenames]{xcolor}
\definecolor{linkcolor}{rgb}{0.0,0.3,0.5}
\usepackage[unicode, colorlinks=true, linkcolor=linkcolor, citecolor=linkcolor, filecolor=linkcolor, urlcolor=linkcolor, linktocpage, breaklinks]{hyperref}
\usepackage[all]{hypcap}
\usepackage{subfigure}
\usepackage[T1]{fontenc}
\usepackage[utf8]{inputenc}
\usepackage[usenames,dvipsnames]{xcolor}
\hypersetup{colorlinks=true,citecolor=romared,linkcolor=romared,urlcolor=romared}

\definecolor{romared}{RGB}{142,0,28}

\newcommand{\be}{\begin{equation}}
\newcommand{\ee}{\end{equation}}

\def\be{\begin{equation}}
\def\ee{\end{equation}}
\newcommand{\beq}{\begin{eqnarray}}
\newcommand{\eeq}{\end{eqnarray}}

\usepackage{aas_macros}
\usepackage{makecell}
\usepackage{soul}
\usepackage{amssymb}
\usepackage{longtable}

\usepackage{lipsum}

\usepackage{orcidlink}
\usepackage{bm}
\let\Re\relax
\DeclareMathOperator{\Re}{Re}

\newcolumntype{Y}{>{\centering\arraybackslash}X}

\newcommand{\aye}{\mathrm{i}}
\newcommand{\e}{\mathrm{e}}
\newcommand{\diff}{\mathrm{d}}

\begin{document}

\title{
Detectability of avoided crossings in black hole ringdowns}

\author{Hayato Imafuku
\orcidlink{0009-0001-3490-8063}}
\email{imafuku-hayato@resceu.s.u-tokyo.ac.jp}
\affiliation{Graduate School of Science, The University of Tokyo, Tokyo 113-0033, Japan}
\affiliation{Research Center for the Early Universe (RESCEU), Graduate School of Science, The University of Tokyo, Tokyo 113-0033, Japan}

\author{Naritaka Oshita
\orcidlink{0000-0002-8799-1382}}
\email{oshita@phys.kindai.ac.jp}
\affiliation{Department of Physics, Kindai University, Osaka 577-8502, Japan}
\affiliation{RIKEN iTHEMS, Wako, Saitama, 351-0198, Japan}

\author{Hiroki Takeda
\orcidlink{0000-0001-9937-2557}}
\email{takeda@tap.scphys.kyoto-u.ac.jp}
\affiliation{The Hakubi Center for Advanced Research, Kyoto University,
Yoshida Ushinomiyacho, Sakyo-ku, Kyoto 606-8501, Japan}
\affiliation{Department of Physics, Kyoto University, Kyoto 606-8502, Japan}

\preprint{RESCEU-14/26}
\preprint{RIKEN-iTHEMS-Report-26}

\begin{abstract}
Quasinormal modes (QNMs) of black holes can exhibit avoided crossings (ACs), in which specific QNM frequencies approach each other while their amplitudes are enhanced and acquire nearly opposite phases, leading to characteristic interference.
Resolving such closely spaced modes through black hole spectroscopy is observationally challenging. 
In this paper, we investigate the detectability of nearly degenerate QNMs in the presence of an AC within a Bayesian framework using three waveform models.
We examine how the inference of the complex frequencies and amplitudes depends on the separation between the two QNM frequencies and on the choice of template waveform. 
We find that resolving the individual QNM frequencies is difficult even under optimistic conditions.
On the other hand, collective waveform signatures associated with ACs may still be identified through complementary waveform descriptions, provided that the AC-related modes dominate the observed ringdown signal and contamination from more slowly damped modes is negligible or can be removed.
\end{abstract}

\maketitle

\section{Introduction}
The ringdown phase of a black hole provides a powerful probe for testing gravity through gravitational-wave (GW) observations.
Owing to the dissipative boundary conditions imposed at the horizon and at infinity, black hole perturbation equations \cite{Regge:1957td,Zerilli:1970se,Teukolsky:1973ha,Sasaki:1981sx,Chandrasekhar:1976zz} can be viewed as non-Hermitian Schr\"{o}dinger-type equations. 
The corresponding complex eigenfrequencies are known as quasinormal-mode (QNM) frequencies (for reviews, see, e.g., \cite{Kokkotas:1999bd, Berti:2009kk, Berti:2025hly}). 
The GW signal in the ringdown phase is described by a superposition of QNMs, together with late-time tail contributions arising from wave backscattering off the spacetime curvature. 

Since the first detection of GWs from a binary black hole merger, GW150914, the LIGO–Virgo–KAGRA observations have provided direct access to the highly dynamical, strong-field regime of gravity \cite{LIGOScientific:2016aoc}. 
The accumulated catalog of compact-binary coalescences \cite{LIGOScientific:2025slb} has enabled increasingly stringent tests of general relativity (GR) \cite{LIGOScientific:2026qni, LIGOScientific:2026fcf, LIGOScientific:2026wpt}. 
In particular, ringdown analyses aim to test the Kerr nature of the remnant black holes \cite{LIGOScientific:2026wpt, LIGOScientific:2025rid, LIGOScientific:2025wao}.
However, resolving multiple QNMs remains challenging with current detector sensitivities and available postmerger signal-to-noise ratios (SNRs).
This observational situation motivates careful assessments of which aspects of the QNM spectrum can be inferred from realistic ringdown data.

The non-Hermitian nature of black hole perturbations implies that phenomena familiar from non-Hermitian systems---such as avoided crossings (ACs) and exceptional points (EPs)---can arise in black hole perturbations \cite{Dias:2021yju,Davey:2022vyx,Dias:2022oqm,Davey:2023fin,Motohashi:2024fwt,Cavalcante:2024swt} (for subsequent works, see, e.g., Refs.~\cite{Oshita:2025ibu,Lo:2025njp,Yang:2025dbn,Takahashi:2025uwo,Kubota:2025hjk,Cao:2025afs,PanossoMacedo:2025xnf,Wu:2025wbp,Nakamoto:2026lyo,Cheng:2026gxu,Zhou:2026xzf}).
An AC refers to the behavior in which two complex eigenfrequencies approach each other and then repel as system parameters are varied.
%
An EP, on the other hand, is a set of parameters for which not only the eigenfrequencies but also the associated eigenfunctions coincide.

A natural question is to what extent such non-Hermitian phenomena can be probed observationally through GWs from black holes. 
In particular, identifying individual QNMs from ringdown signals via black hole spectroscopy appears to be especially challenging in this context. 
This difficulty arises because, near an AC (while an exact EP would generally require fine-tuning of the system parameters), the relevant QNM frequencies become very close to each other, making them difficult to resolve observationally.
Although the individual amplitudes of the two QNMs are enhanced \cite{Motohashi:2024fwt}, it has been analytically shown that \cite{Oshita:2025ibu} the corresponding modes are excited with nearly opposite signs, leading to destructive interference.
As a characteristic signature of QNMs near an AC, their interference can produce a linear-growth term proportional to $t \e^{-\aye\omega t}$ in the time-domain waveform \cite{Yang:2025dbn}, which can become prominent under certain conditions \cite{Nakamoto:2026lyo}.
Previous works based on noiseless waveforms \cite{Yang:2025dbn,PanossoMacedo:2025xnf} have shown that a model approximating the signal as a single double-pole QNM provides a better description than a simple superposition of damped sinusoids.

In this paper, we investigate observational signatures of ACs in black hole ringdown signals.
Specifically, we explore the resolvability of nearly degenerate QNM frequencies and the measurability of characteristic waveform amplitudes using Bayesian parameter inference.
We also employ three waveform models: a simple superposition of two damped sinusoids, an AC-inspired model incorporating the characteristic destructive interference between nearly degenerate modes, and an effective EP model in which the two QNMs are approximated by a double-pole QNM.
Our analysis suggests that, even when individual QNMs are difficult to resolve near ACs, EP-inspired waveform descriptions may capture characteristic signatures of the signal, at least when (i) the least-damped mode experiences an AC and (ii) the excitation of modes unrelated to the AC is negligible.

This paper is organized as follows. 
In Sec.~\ref{sec:Theory: a review}, we provide a brief review of black hole perturbation theory, as well as ACs, EPs, and their unique feature in QNM excitation. 
In Sec.~\ref{sec:Data analysis}, we investigate the detectability and separability of two QNMs relevant to an AC using three different waveform models. 
In Sec.~\ref{sec:Discussion_Conclusion}, our discussion and conclusions are provided.

Throughout this paper, we adopt geometric units, setting $G=c=1$.
\section{Formalism}
\label{sec:Theory: a review}

\subsection{Excitation of QNMs}
The radial perturbation of a test field around a Schwarzschild or Kerr black hole is governed by the Schr\"{o}dinger-type wave equation in the frequency space,
\begin{equation}
    \left[ \frac{\diff^2}{\diff x^2} - V_{\ell m \omega}(r) \right] \psi_{\ell m \omega} (x) = {\cal T}_{\ell m \omega}(x)\,,
    \label{wave_eq_for_BH_pert}
\end{equation}
where $x$ is the tortoise coordinate, $V_{\ell m \omega}(r)$ is the potential barrier, ${\cal T}_{\ell m \omega}(x)$ is the source term, and $(\ell, m)$ is the angular modes for the spherical/spheroidal perturbation decomposition.
In the case of Kerr background with the spin parameter $a$ and mass $M$, we can set the perturbation variable $\psi_{\ell m \omega} (x)$ so that the potential barrier becomes short range, i.e., $V = - k_{\rm H}^2 + {\cal O}(e^{2\kappa x})$ with $k_{\rm H} \coloneqq \omega - m a / (2 M r_+)$ at $x \to - \infty$, and $V = -\omega^2 + {\cal O}(r^{-2})$ at $r \to \infty$ \cite{Chandrasekhar:1976zz,Sasaki:1981sx}.
For such a perturbation variable, we can impose the free-oscillation boundary condition as
\begin{align}
    \psi_{\ell m \omega} \to
    \begin{cases}
        \e^{-\aye k_{\rm H} x} \ &(x \to -\infty)\,,\\
        \e^{+\aye \omega x} \ &(x \to +\infty)\,.
    \end{cases}
\end{align}

One can solve the wave equation \eqref{wave_eq_for_BH_pert} by utilizing the Green's function technique:
\begin{equation}
    \psi_{\ell m \omega} (x) = \int \diff x' G_{\ell m \omega}(x, x') {\cal T}_{\ell m \omega} (x')\,,
\end{equation}
where $G_{\ell m \omega}(x, x')$ is the Green's function associated with the free-oscillation boundary condition.
The Green's function is constructed as
\begin{align}
G_{\ell m \omega} (x, x') = \frac{-1}{2 \aye \omega A_{\rm in} (\omega)}
\begin{cases}
    \psi^{\rm in}_{\ell m \omega} (x) \psi^{\rm up}_{\ell m \omega} (x')\,, \ (x < x')\,,\\
    \psi^{\rm in}_{\ell m \omega} (x') \psi^{\rm up}_{\ell m \omega} (x)\,, \ (x > x')\,,
\end{cases}
\end{align}
where $\psi^{\rm in}_{\ell m \omega} (x)$ and $\psi^{\rm up}_{\ell m \omega} (x)$ are two independent homogeneous solutions to \eqref{wave_eq_for_BH_pert}, which satisfy the following boundary condition:
\begin{align}
    \psi^{\rm in}_{\ell m \omega} (x) &=
\begin{cases}
    \e^{-\aye k_{\rm H} x}\,, \ &(x \to - \infty)\,,\\
    A_{\rm out} \e^{+\aye \omega x} + A_{\rm in} \e^{-\aye \omega x}\,, \ &(x \to + \infty)\,,
\end{cases}\\
\psi^{\rm up}_{\ell m \omega} (x) &=
\begin{cases}
    B_{\rm out} \e^{\aye k_{\rm H} x} + B_{\rm in} \e^{-\aye k_{\rm H} x}\,, \ &(x \to - \infty)\,,\\
    \e^{+\aye \omega x}\,, \ &(x \to + \infty)\,.
\end{cases}
\end{align}
The Green's function has poles at $\omega = \omega_{\ell m n}^{\pm} \in {\mathbb C}$, which are known as the QNM frequencies.
The upper index $+ \ (-)$ in $\omega_{\ell m n}^{\pm}$ stands for the prograde (retrograde) QNM frequency, whose real part is positive (negative).
The QNM poles appear in the lower-half region in the complex-$\omega$ plane, which leads to the damped oscillation in the time-domain signal of the black hole ringing,
\begin{equation}
    \Psi_{\ell m} (t,x) = \frac{1}{2 \pi} \int_{\cal C} \diff \omega \psi_{\ell m \omega} (x) \e^{-\aye \omega t}\,, 
\end{equation}
as one can deform the contour ${\cal C}$ in the lower-half region \cite{Leaver:1986gd}.
For $x \to \infty$, the QNM excitation in the ringdown waveform $h^{\rm (q)}_{\ell m}$ is represented as

\begin{align}
    h^{\rm (q)}_{\ell m} (u) &= \sum_n \frac{\e^{-\aye \omega_{\ell m n} u} A_{\rm out}(\omega_{\ell m n})}{2 \omega_{\ell m n} \alpha_{\ell m n}}\nonumber \\
    & \qquad\times \left. \int \diff x' \frac{\psi^{\rm in}_{\ell m \omega} (x')}{A_{\rm out} (\omega)} {\cal T}_{\ell m \omega}(x')\right|_{\omega = \omega_{\ell m n}}\,,\\
    & = \sum_n E_{\ell m n} T_{\ell m n} \e^{-\aye \omega_{\ell m n} u}\,,
\end{align}
where $u \coloneqq t-x$, $\alpha_{\ell m n} \coloneqq (\diff A_{\rm in} / \diff\omega)_{\omega = \omega_{\ell m n}}$, $E_{\ell m n}$ is the excitation factor \cite{Leaver:1986gd,Sun:1988tz,Andersson:1995zk,Glampedakis:2001js,Glampedakis:2003dn,Berti:2006wq,Zhang:2013ksa}
\begin{equation}
    E_{\ell m n} = \frac{A_{\rm out}(\omega_{\ell m n})}{2 \omega_{\ell m n} \alpha_{\ell m n}}\,,
\end{equation}
and $T_{\ell m n}$ is the source factor
\begin{align}
    T_{\ell m n} &\coloneqq T_{\ell m} (\omega_{\ell m n}) \,,\\
    T_{\ell m} (\omega) &= \int \diff x' \frac{\psi^{\rm in}_{\ell m \omega} (x')}{A_{\rm out} (\omega)} {\cal T}_{\ell m \omega}(x')\,.
\end{align}
Therefore, the amplitude of each QNM, i.e., excitation coefficient $C_{\ell mn}$, is obtained by the product of the excitation factor and the source factor as
\begin{equation}
    C_{\ell mn} = E_{\ell mn} \times T_{\ell mn}\,.
\end{equation}
In the following, we omit the subscripts $\ell$ and $m$ for brevity, but keep the label of the overtone number $n$.

\subsection{QNM excitation in the presence of ACs}
When the Green's function has simple poles and two QNM frequencies, $\omega_i$ and $\omega_j$, are close to each other, i.e.,
\begin{equation}
    |(\omega_i - \omega_j) / \omega_i| \ll 1\,,
\end{equation}
one can describe $A_{\rm in} (\omega)$ around $\omega = \omega_{i}$ as
\begin{equation}
A_{\rm in} (\omega) = \beta^{-1}(\omega) (\omega - \omega_i) (\omega - \omega_j)\,,
\end{equation}
where $\beta(\omega)$ is a regular function around $\omega = \omega_{i}$.
In this case, we have
\begin{align}
    \alpha_i &= \beta^{-1} (\omega_i) \Delta \omega\,,\\
    \alpha_j &= -\beta^{-1} (\omega_j) \Delta \omega\,,
\end{align}
where $\Delta \omega \coloneqq \omega_i - \omega_j$.
Then the QNM excitation coefficients for the $i$ th and $j$ th QNMs take the form of
\begin{align}
    C_i &= E_i T_i = \frac{A_{\rm out} (\omega_i) \beta (\omega_i)}{2 \omega_i \Delta \omega} T_i = \frac{D(\omega_i)}{\Delta \omega}\,,\\
    C_j &= E_j T_j = -\frac{A_{\rm out} (\omega_j) \beta (\omega_j)}{2 \omega_j \Delta \omega} T_j = - \frac{D(\omega_j)}{\Delta \omega}\,,
\end{align}
where
\begin{equation}
D(\omega) \coloneqq \frac{A_{\rm out} (\omega) \beta (\omega) T(\omega)}{2 \omega}\,.
\end{equation}
The superposition of the two QNMs in the time domain $h(t)$ is
\begin{widetext}
\begin{subequations}
\begin{align}
    h(t) &= C_i \e^{-\aye \omega_i t} + C_j \e^{-\aye \omega_j t}\,,
    \label{two_dam_review}\\
    &= \frac{D_i}{\Delta \omega} \e^{-\aye \omega_i t} - \frac{D_i \left(1 - \Delta \omega \left. \frac{\diff \log D}{\diff \omega} \right|_{\omega = \omega_i} + {\cal O}(\Delta \omega^2)\right)}{\Delta \omega} \e^{-\aye \omega_j t}\,,
   \label{ac_approx}
    \\
    & = \left[ \frac{\diff D}{\diff \omega} - \aye t D(\omega) \right]_{\omega = \omega_i} \e^{-\aye \omega_i t} + {\cal O}(\Delta \omega)\,,
    \label{ep_approx}
\end{align}
\end{subequations}
\end{widetext}
where $D_i \coloneqq D(\omega_i)$. 
Equation~\eqref{ac_approx} approximates the excitation coefficient $C_j$ only, and in Eq.~\eqref{ep_approx}, both the amplitude and QNM eigenfunction, $C_j \e^{-\aye \omega_j t}$, are expanded around $\omega = \omega_i$.
In the former approximation, Eq.~\eqref{ac_approx}, the destructive amplification of the two QNMs, whose amplitudes are proportional to $\Delta \omega^{-1}$ \cite{Oshita:2025ibu}, is apparent.
On the other hand, in the latter approximation, Eq.~\eqref{ep_approx}, it describes the excitation of an effective double-pole QNM at the EP that exhibits the linear growth at early times \cite{Yang:2025dbn}.

To verify the AC of QNMs, it is important to perform black hole spectroscopy (or QNM decomposition) for two QNMs exhibiting an AC. 
However, this task becomes increasingly challenging as the frequencies approach each other.
There are various ways to model such QNM excitations. 
Beyond a simple superposition of two damped sinusoids, for example, the waveform can be approximated by \eqref{ac_approx} or \eqref{ep_approx}.
In particular, the latter model \eqref{ep_approx} has been used in the fit analysis \cite{Yang:2025dbn,PanossoMacedo:2025xnf}, and it has been shown to work well when the two QNM frequencies are very close.
However, the model reduces the number of model parameters and treats the two QNMs as a double-pole QNM.
In the next section, we study the detectability or resolvability of two QNMs undergoing the AC.
We then discuss which QNM model is efficient to characterize the (destructively enhanced) amplitude of the two QNMs.

\section{Data analysis}
\label{sec:Data analysis}
As discussed in Sec.~\ref{sec:Theory: a review}, QNMs associated with ACs can have closely spaced complex frequencies and enhanced amplitudes with nearly opposite phases.
In the EP limit, the superpositions of such modes can be described effectively by a double-pole waveform with a linearly growing component.
In this section, we investigate the detectability of these AC-related waveform signatures using three waveform models within a Bayesian framework. 
Specifically, we consider idealized two-mode ringdown signals and vary the separation between the two complex frequencies. 
This allows us to examine the resolvability of the complex frequencies and the measurability of the characteristic amplitude parameters, as well as their dependence on the choice of template waveform. 
Our analysis includes simulated detector noise and performs Bayesian parameter estimation in the time domain. 
Although the setup is intentionally optimistic, it provides a controlled benchmark for assessing the fundamental detectability of AC-related signatures.

Section~\ref{subsec:Waveform model} introduces the three waveform models used as templates in our analysis, and Sec.~\ref{subsec:Bayesian framework} summarizes the Bayesian framework and analysis settings. 
The results are presented in Secs.~\ref{subsubsec:Agnostic injection with two damped sinusoids} and \ref{subsubsec:Application to avoided crossing in GR}. 
In the former, we analyze controlled injection signals with systematically varied separations in the complex-frequency plane, while in the latter, we apply the same analysis to an example of an AC in GR.


\subsection{Waveform model}
\label{subsec:Waveform model}
The GW strain observed by the $I$ th detector is written as
\begin{align}
    h_{I}(t) = \Re [F^I h(t)] \:,
\end{align}
where $F^I=F^{I}_{+}+\aye F^{I}_{\times}$ is the complex antenna pattern function of the $I$ th detector, constructed from the plus ($+$) and cross ($\times$) polarization responses. 
The antenna pattern is specified by the geometric parameters: the right ascension ($\mathrm{RA}$), the declination ($\mathrm{Dec}$), the polarization angle $\psi$, and the coalescence time $t_c$.
The complex GW waveform is defined as $h(t)=h_{+}(t)-\aye h_{\times}(t)$, which corresponds to the ringdown waveform model used in this study. 
We consider the following three models to investigate the detectability and resolvability of two QNMs with different separations in frequency and damping time.

\subsubsection{2DS model}
\label{subsubsec:2DS model}
The 2DS model consists of a simple superposition of two damped sinusoids, as shown in Eq.~\eqref{two_dam_review}, given by
\begin{align}
    h(t) &= A_1 \e^{-t/\tau_1} \e^{\aye\left(2\pi f_1t + \phi_{1}\right)} \nonumber \\
    &\quad + A_2 \e^{-t/\tau_2} \e^{\aye\left(2\pi f_2t + \phi_{2}\right)} \:,
\end{align}
where $A_i$, $f_i$, $\tau_i$, and $\phi_i$ denote the real-valued amplitude, frequency, damping time, and initial phase of the $i$ th QNM $(i\in \{1,2\})$, respectively. 
The waveform convention is chosen to be consistent with the standard Fourier-transform convention commonly adopted in GW data analysis, although the inference itself is performed in the time domain.
We also require the damping times to be positive.

\subsubsection{2AC model}
\label{subsubsec:2AC model}
The 2AC model also consists of two damped modes, but incorporates the characteristic enhancement and the destructive interference associated with an AC. 
This model corresponds to Eq.~\eqref{ac_approx} and is written as
\begin{align}
    h(t) &= \frac{A}{\delta \omega} \e^{-t/\tau_1} \e^{\aye(2\pi f_1 t + \phi_{A})} \nonumber \\
    &\quad -\frac{A(1 + \alpha \e^{\aye\phi_{\alpha}} \delta \omega)}{\delta \omega} \e^{-t/\tau_2} \e^{\aye(2\pi f_2 t + \phi_{A})} \,,\label{eq:2AC}
\end{align}
where $A$, $\alpha$, $\phi_A$, and $\phi_{\alpha}$ are real-valued model parameters.
Here, $\delta \omega \coloneqq \omega_1 - \omega_2$ is the separation between the complex frequencies, defined by $\omega_i=2\pi f_i-\aye/\tau_i$. 
The complex parameters $A \e^{\aye \phi_A}$ and $\alpha \e^{\aye \phi_{\alpha}}$ correspond to $D (\omega_1)$ and $-\diff \log D/\diff \omega|_{\omega = \omega_1}$, respectively [cf.~Eq.~\eqref{ac_approx}]. 
As shown later, this waveform parametrization is useful for characterizing the quantities $D$ and $\diff D/\diff \omega$, which determine when the linear-growth component induced by nearly degenerate QNM interference becomes significant in the presence of an AC \cite{Nakamoto:2026lyo}. 

\subsubsection{EP model}
\label{subsubsec:EP model}
The 2AC model can be further approximated by a single effective double-pole QNM with frequency $f$ and damping time $\tau$. 
This approximation defines the effective EP model described by Eq.~\eqref{ep_approx}.
The EP waveform introduced in Eq.~\eqref{ep_approx} is reparametrized as
\begin{align}
    h(t) = \left(C\e^{\aye\phi_C} + \aye D\e^{\aye\phi_D}t\right) \e^{-t/\tau} \e^{\aye 2\pi ft} \:, \label{eq:EP}
\end{align}
where $C$ and $\phi_C$ denote the amplitude and phase of the damped-sinusoid component, respectively, while $D$ and $\phi_D$ characterize the amplitude and phase of the linearly growing component. 
The correspondence to the 2AC model can be obtained either by comparing Eqs.~\eqref{ac_approx} and \eqref{ep_approx} or, equivalently, by taking the limit $\delta\omega \to 0$ in Eq.~\eqref{eq:2AC}, which yields $C=A\alpha$, $D=A$, $f=f_1$, $\tau=\tau_1$, $\phi_C=\phi_A+\phi_\alpha+\pi$, and $\phi_D=\phi_A$.

\medskip
Note that the 2AC model is closely related to the 2DS waveform as an AC-adapted parametrization in the nearly degenerate regime, while the EP model corresponds to its effective double-pole limit.
For the 2DS and 2AC models, the corresponding posterior measures can, in principle, be related by an appropriate parameter transformation and prior reweighting when the approximation remains valid.
The EP model, by contrast, provides a reduced effective description that makes the linear-growth component explicit.
Thus, the three descriptions are not independent physical waveform families, but they emphasize different aspects of the same nearly degenerate QNM structure.

\subsection{Bayesian framework}
\label{subsec:Bayesian framework}
Parameter estimation is performed using Bayesian inference, which evaluates the posterior probability distribution $p(\bm{\theta}|\bm{d}, M)$ for the parameter set $\bm{\theta}$, given the data $\bm{d}$ and a waveform model $M$. 
According to Bayes' theorem~\cite{Bayes1763, Maggiore:2007ulw, LIGOScientific:2019hgc},
\begin{align}
    p(\bm{\theta}|\bm{d},M) = \frac{p(\bm{\theta}|M)p(\bm{d}|\bm{\theta},M)}{p(\bm{d}|M)}\:,
\end{align}
where $p(\bm{\theta}|M)$ is the prior probability distribution, $p(\bm{d}|\bm{\theta},M)$ is the likelihood, and $p(\bm{d}|M)$ is the evidence.
Here, $\bm{d}=\{d_I\}^{N}_{I=1}$ denotes the strain data from the network of $N$ detectors.
The waveform model $M$ corresponds to one of the three ringdown waveform models introduced in Sec.~\ref{subsec:Waveform model}. 
The full parameter set is written as $\bm{\theta} = \bm{\theta}_{\mathrm{m}} \cup \bm{\theta}_{\mathrm{g}}$, where $\bm{\theta}_{\mathrm{m}}$ with $\mathrm{m}\in\{\mathrm{2DS},\mathrm{2AC},\mathrm{EP}\}$ denotes the waveform-model parameters and $\bm{\theta}_{\mathrm{g}}=\{\mathrm{RA},\mathrm{Dec},\psi,t_c\}$ contains the geometric parameters specifying the antenna responses. 
The waveform-model parameter sets are
\begin{align}
    \bm{\theta}_{\mathrm{2DS}} &= \{A_1, A_2, f_1, f_2, \tau_1, \tau_2, \phi_1, \phi_2\} \:, \\
    \bm{\theta}_{\mathrm{2AC}} &= \{A, \alpha, f_1, f_2, \tau_1, \tau_2, \phi_A, \phi_{\alpha}\} \:, \\
    \bm{\theta}_{\mathrm{EP}} &= \{C, D, f, \tau, \phi_C, \phi_D\} \:.
\end{align}

The prior distributions $p(\bm{\theta}|M)$ for the model parameters $\bm{\theta}_{\mathrm{m}}$ are chosen to be uniform over the linear ranges specified below.
The prior ranges are $[0,10^{-17}]$ for $A_i$, $[0,10^{-15}]~\mathrm{s}^{-1}$ for $A$, $[0,10^{-16}]$ for $C$, $[0,10^{-16}]~\mathrm{s}^{-1}$ for $D$, $[0,1]~\mathrm{s}$ for $\alpha$, $[21,500]~\mathrm{Hz}$ for the frequencies, $[5\times10^{-4},5\times10^{-2}]~\mathrm{s}$ for the damping times, and $[-\pi,\pi]$ for the phases. 
We have verified that the final results are insensitive to the precise choice of these prior ranges. 
To avoid label switching between the two modes in the 2DS and 2AC models, we impose the ordering condition, either $\tau_1>\tau_2$ or $f_1>f_2$. 
For the geometric parameters $\bm{\theta}_{\mathrm{g}}$, $\delta$-function priors centered on the injected values are adopted, effectively fixing these parameters.
Consequently, only the waveform-model parameters $\bm{\theta}_{\mathrm{m}}$ are sampled in the present analysis.

The likelihood $p(\bm{d}|\bm{\theta},M)$ is computed in the time domain, assuming stationary Gaussian detector noise. 
Given the strain data $d_{I}(t)=h_{I}(t)+n_{I}(t)$, the likelihood is written as
\begin{align}
    p(\bm{d}|\bm{\theta},M) &\propto \exp\Big[-\frac{1}{2}\sum_{I} \int \diff t \int \diff t' \ (d_{I}(t)-h_{I}(t)) \nonumber \\
    &\quad \times C^{-1}_{I}(t-t') (d_{I}(t')-h_{I}(t'))\Big] \:.
\end{align}
Here $C_I^{-1}(t-t')$ denotes the inverse covariance kernel associated with the noise autocovariance function $C_I(t-t')$,
\begin{align}
    C_I(t-t')
    =
    \left\langle
    n_I(t)n_I(t')
    \right\rangle
    \:,
\end{align}
which is determined by the detector power spectral density.
In this study, we consider a three-detector network consisting of LIGO Hanford, LIGO Livingston, and Virgo.
The noise $n_{I}(t)$ is generated as colored Gaussian noise based on the power spectral density of each detector at the O4 design sensitivity~\cite{Abbott2020Prospects}.

Our Bayesian analysis is performed using the \texttt{pyRing} package~\cite{Carullo:2019flw, Isi:2019aib, LIGOScientific:2020tif, pyRing}, employing the \texttt{CPNest} sampler~\cite{john_veitch_2025_17334573} for nested sampling. 
We set both the number of live points (\texttt{nlive}) and the maximum Markov chain Monte Carlo chain length used for proposing new samples (\texttt{maxmcmc}) to 2048. 
We have verified that these settings are sufficient to ensure sampling convergence.

\subsection{Results}
\label{subsec:Results}
\subsubsection{Controlled injections of nearly degenerate QNM pairs}
\label{subsubsec:Agnostic injection with two damped sinusoids}
To evaluate the detectability of AC signatures in ringdown signals, we consider controlled injections of nearly degenerate QNM pairs whose complex-frequency separations are systematically varied.
In realistic situations, not only the two QNMs associated with the AC but also the excitation of other modes irrelevant to the AC is expected to contribute. Here, however, as an optimistic scenario, we assume that the QNMs approaching each other due to the AC are (i) the least-damped modes and (ii) sufficiently dominant that the excitation of other modes irrelevant to the AC can be neglected.
We then discuss the best-case detectability and separability of the two QNMs.

We consider various separations between the two QNM frequencies.
We separately analyze the cases with shifts only in the frequency (frequency-shifted injections), i.e., $f_1 > f_2$ and $\tau_1 = \tau_2$, and those with shifts only in the damping time (damping-time-shifted injections), i.e., $f_1 = f_2$ and $\tau_1 > \tau_2$. 
The fractional complex-frequency separation between the two modes is quantified by $|\delta\omega/\omega_1|$, and we consider the cases $|\delta\omega/\omega_1| = 0.3, 0.1, 0.01$, and $0.001$.
The injections are constructed using the 2AC model. 
Since the 2AC and 2DS models are related by a reparametrization, this setup is equivalent to constructing the injections from the 2DS model, up to the chosen parameter ranges.

For the frequency-shifted injections, only the frequency of the second mode, $f_2$, is varied, while $f_1$, $\tau_1$, and $\tau_2$ are fixed for all values of $|\delta\omega/\omega_1|$.
We set $f_1 \sim 201.2~\mathrm{Hz}$ and $\tau_1 = \tau_2 \sim 3.322~\mathrm{ms}$, corresponding to the complex frequency of the $(2,2,0)$ mode of a Schwarzschild black hole with mass $M=60~M_{\odot}$.
The corresponding values of the second-mode frequency are $f_2 \sim 139.2, 180.6, 199.2$, and $201.0~\mathrm{Hz}$ for $|\delta\omega/\omega_1|=0.3$, $0.1$, $0.01$, and $0.001$, respectively. 
In all cases, the amplitude parameter $A$ is chosen such that the SNR is $\sim 100$.
Since there is no prior knowledge of the source term, we set $\alpha = M$ based on dimensional considerations, yielding $\alpha \sim 0.296~\mathrm{ms}$ for $M=60~M_{\odot}$. 
The phases $\phi_A$ and $\phi_{\alpha}$ are set to $0~\mathrm{rad}$, and all geometrical parameters $\bm{\theta}_{\mathrm{g}}$ are fixed to $0$.

For the damping-time-shifted injections, $\tau_2$ is determined based on the values of $\tau_1$ and $|\delta\omega/\omega_1|$ using the same procedure as in the frequency-shifted cases. 
We set $f_1=f_2\sim201.2~\mathrm{Hz}$ and $\tau_1\sim3.322~\mathrm{ms}$, yielding $\tau_2\sim1.447, 2.320, 3.184$, and $3.308~\mathrm{ms}$ for $|\delta\omega/\omega_1|=0.3, 0.1, 0.01$, and $0.001$, respectively. 
The remaining parameters are chosen identically to those of the frequency-shifted injections.

In total, we analyze eight injection configurations spanning a range of complex-frequency separations using the three waveform models introduced in Sec.~\ref{subsec:Waveform model}.
In addition, we also verify that the qualitative trends discussed below remain unchanged when both the frequencies and damping times are varied simultaneously or when different SNRs are adopted.
As expected, the posterior uncertainties decrease with increasing SNR, while the overall inference behavior remains qualitatively similar.

\begin{figure*}[t]
    \begin{tabular}{cc}
      \begin{minipage}[t]{0.45\textwidth}
        \centering
        \includegraphics[keepaspectratio, width=0.9\linewidth]{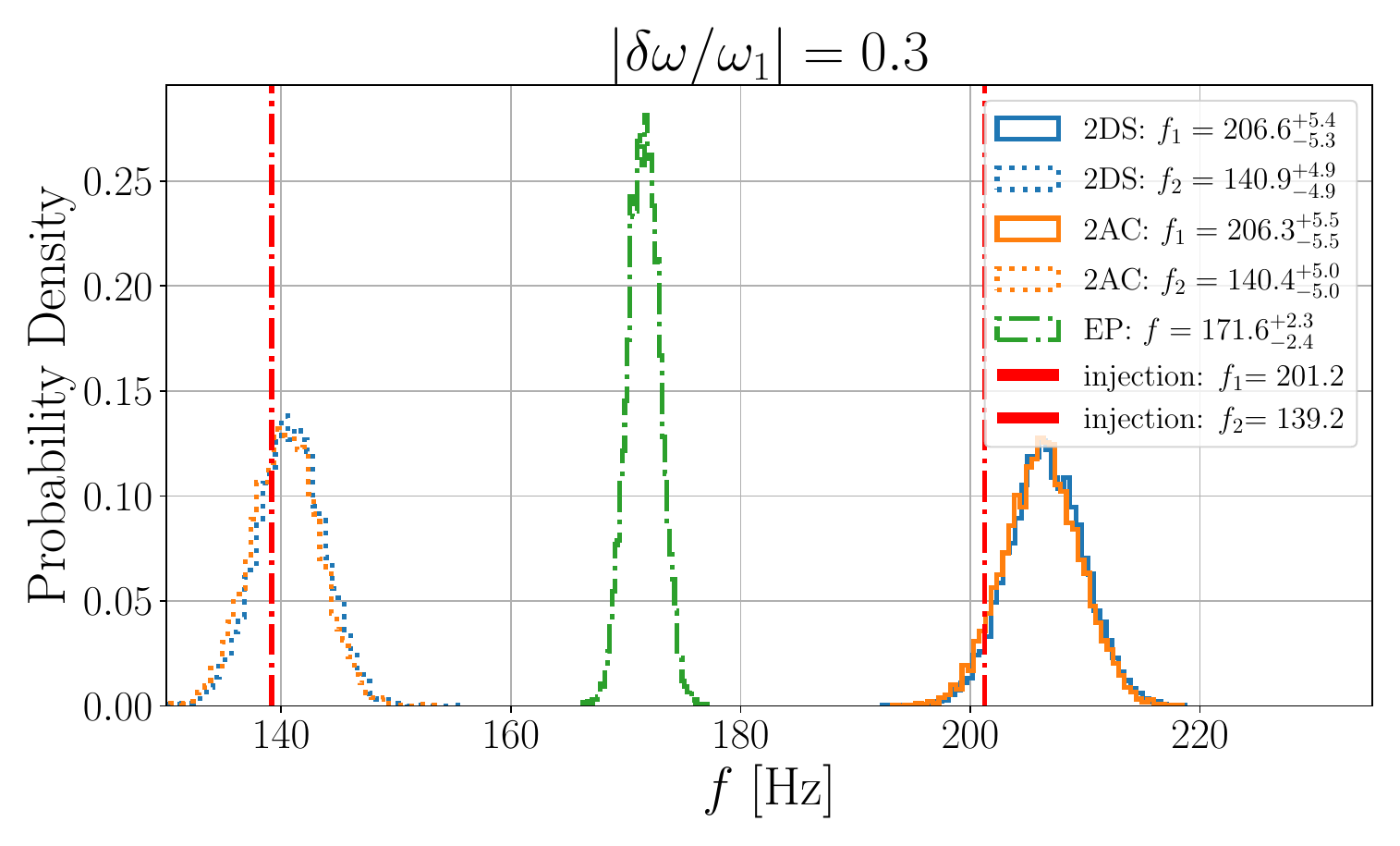}
      \end{minipage} &
      \begin{minipage}[t]{0.45\textwidth}
        \centering
        \includegraphics[keepaspectratio, width=0.9\linewidth]{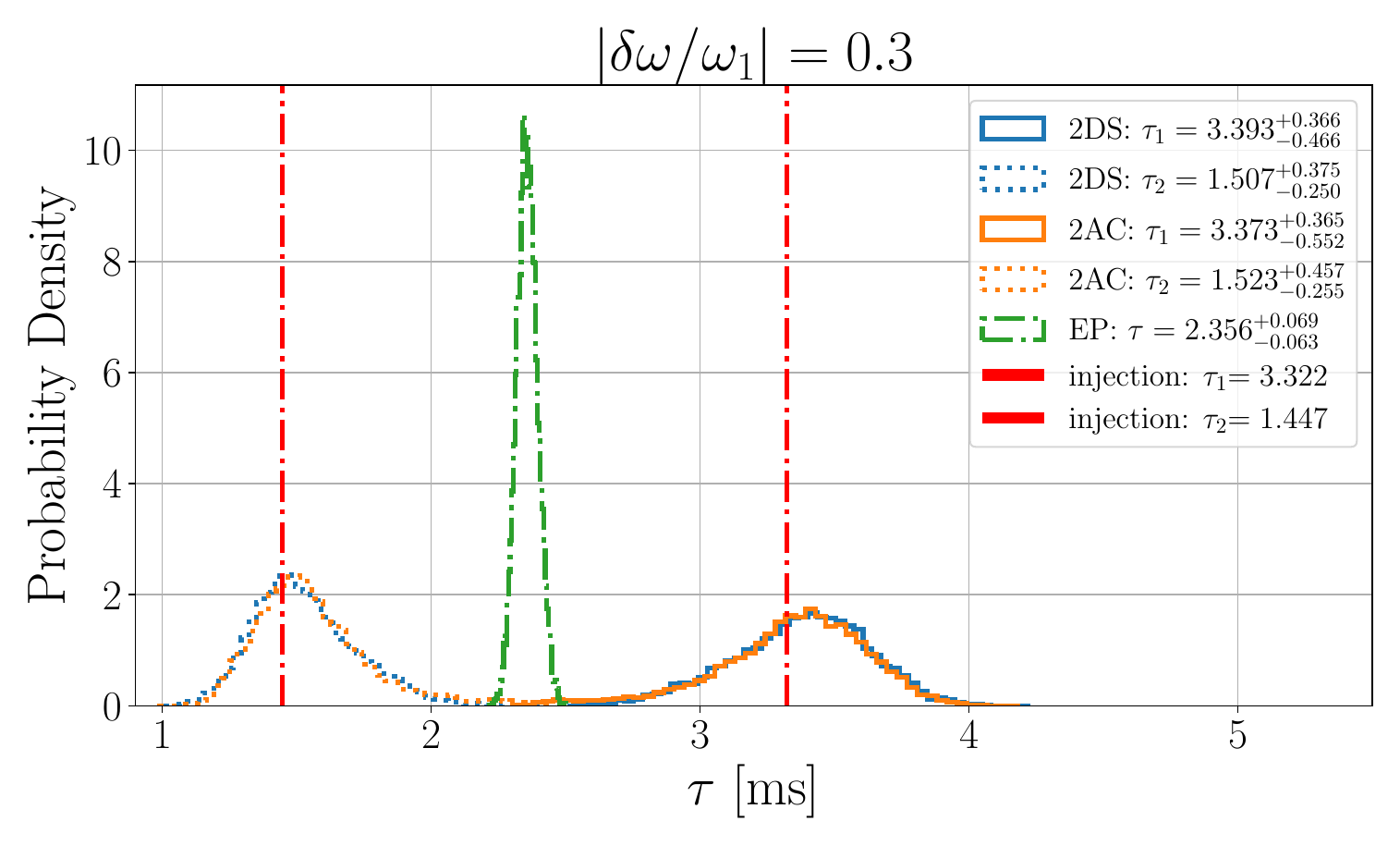}
      \end{minipage} \\
      \begin{minipage}[t]{0.45\textwidth}
        \centering
        \includegraphics[keepaspectratio, width=0.9\linewidth]{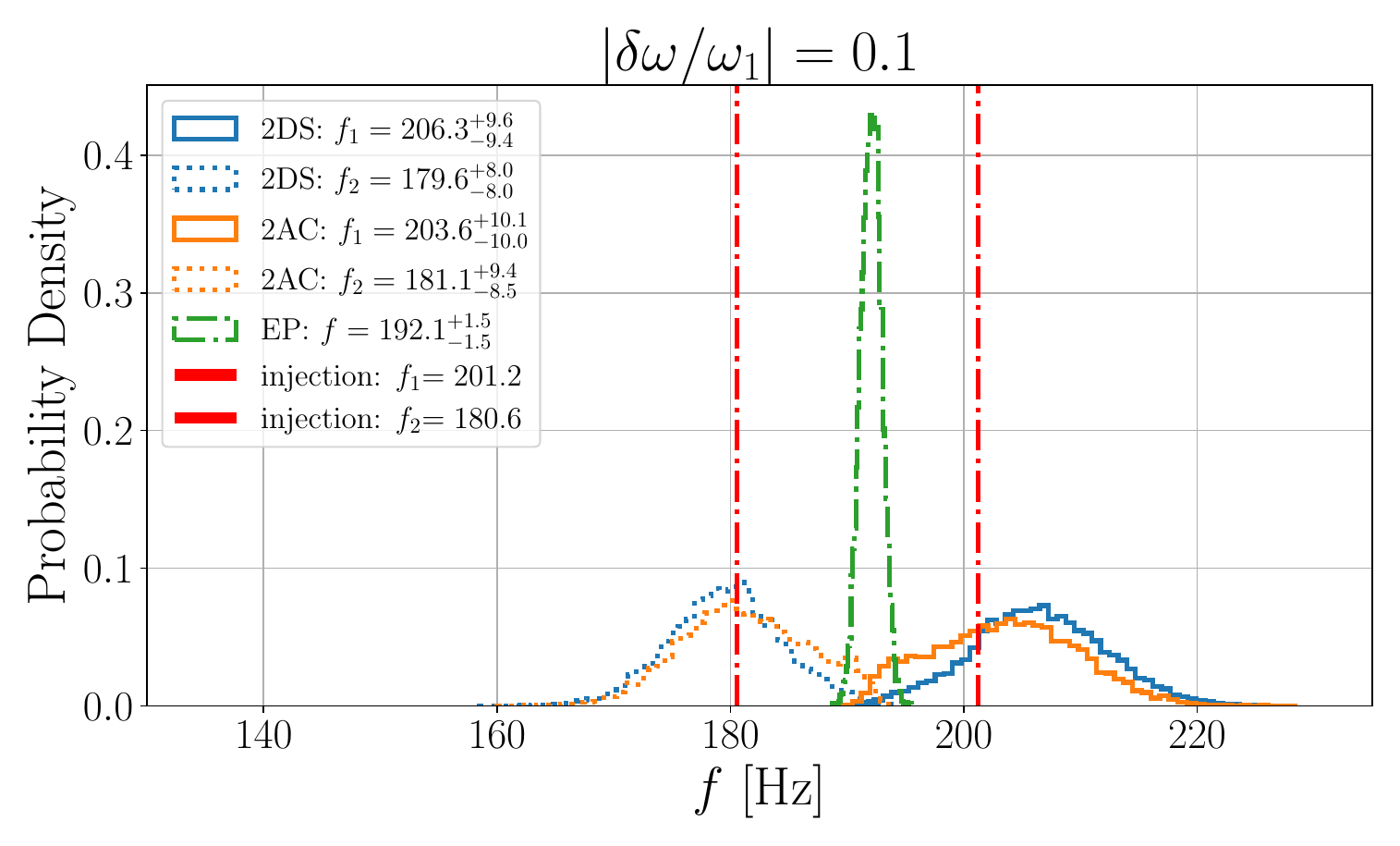}
      \end{minipage} &
      \begin{minipage}[t]{0.45\textwidth}
        \centering
        \includegraphics[keepaspectratio, width=0.9\linewidth]{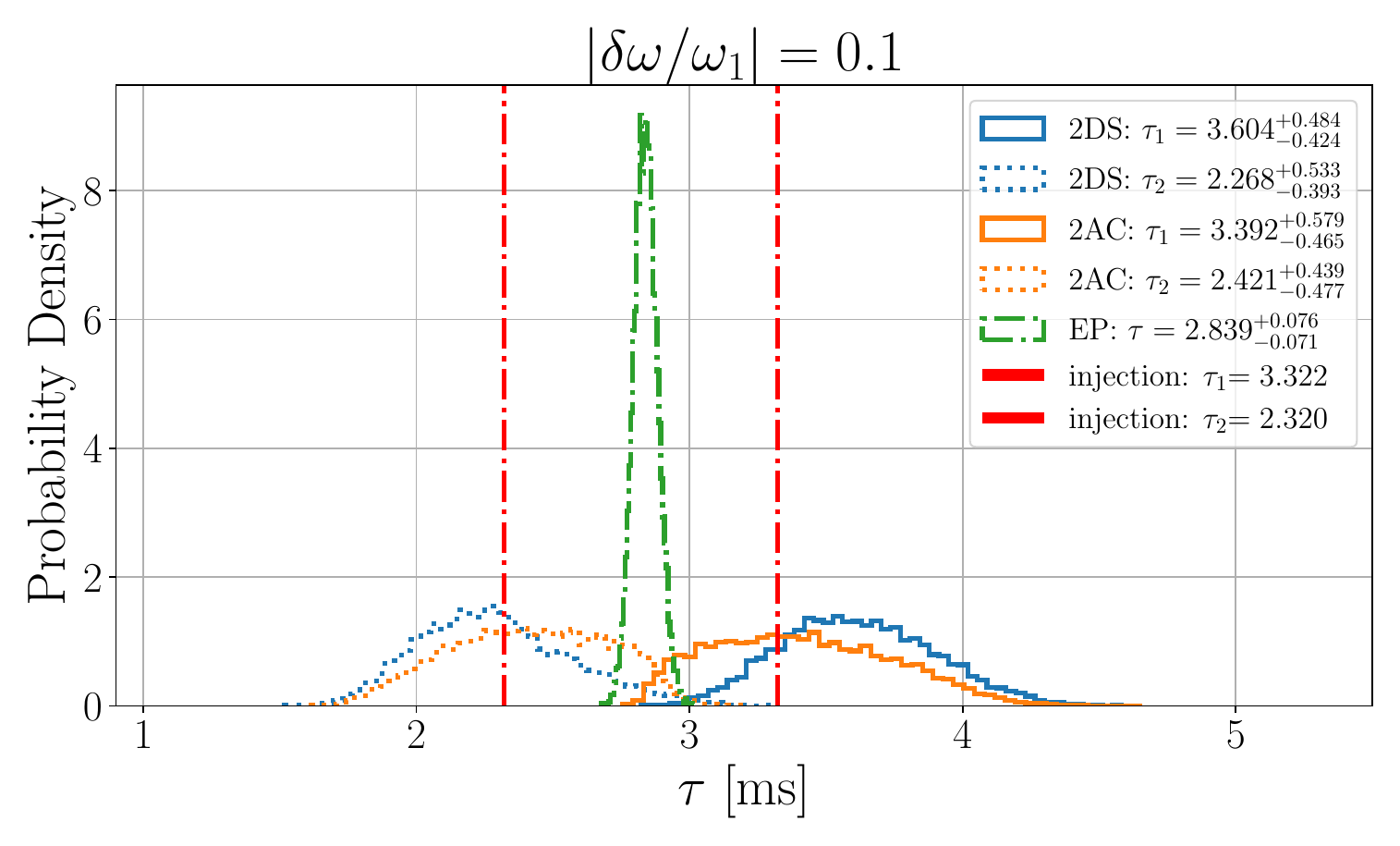}
      \end{minipage} \\
      \begin{minipage}[t]{0.45\textwidth}
        \centering
        \includegraphics[keepaspectratio, width=0.9\linewidth]{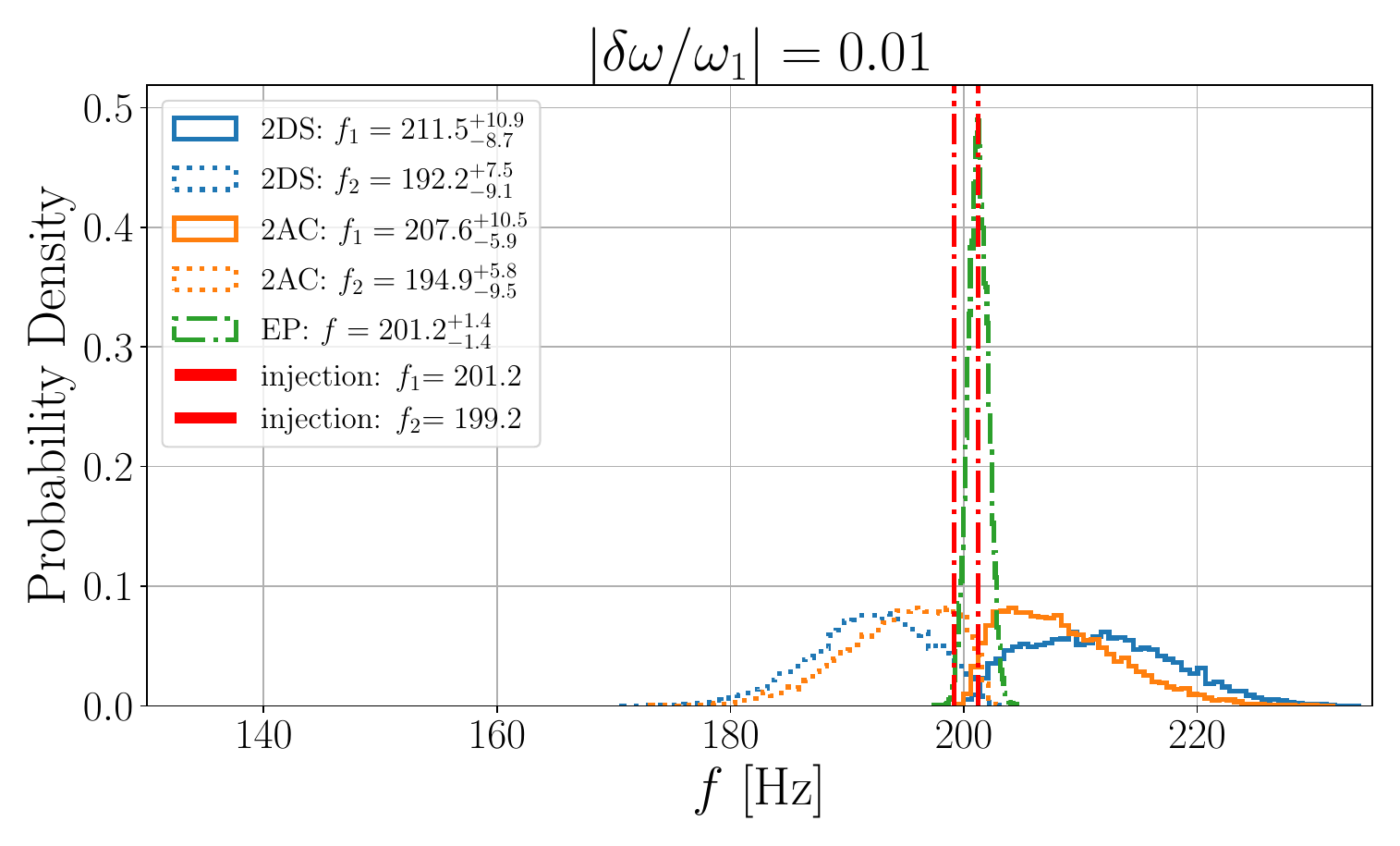}
      \end{minipage} &
      \begin{minipage}[t]{0.45\textwidth}
        \centering
        \includegraphics[keepaspectratio, width=0.9\linewidth]{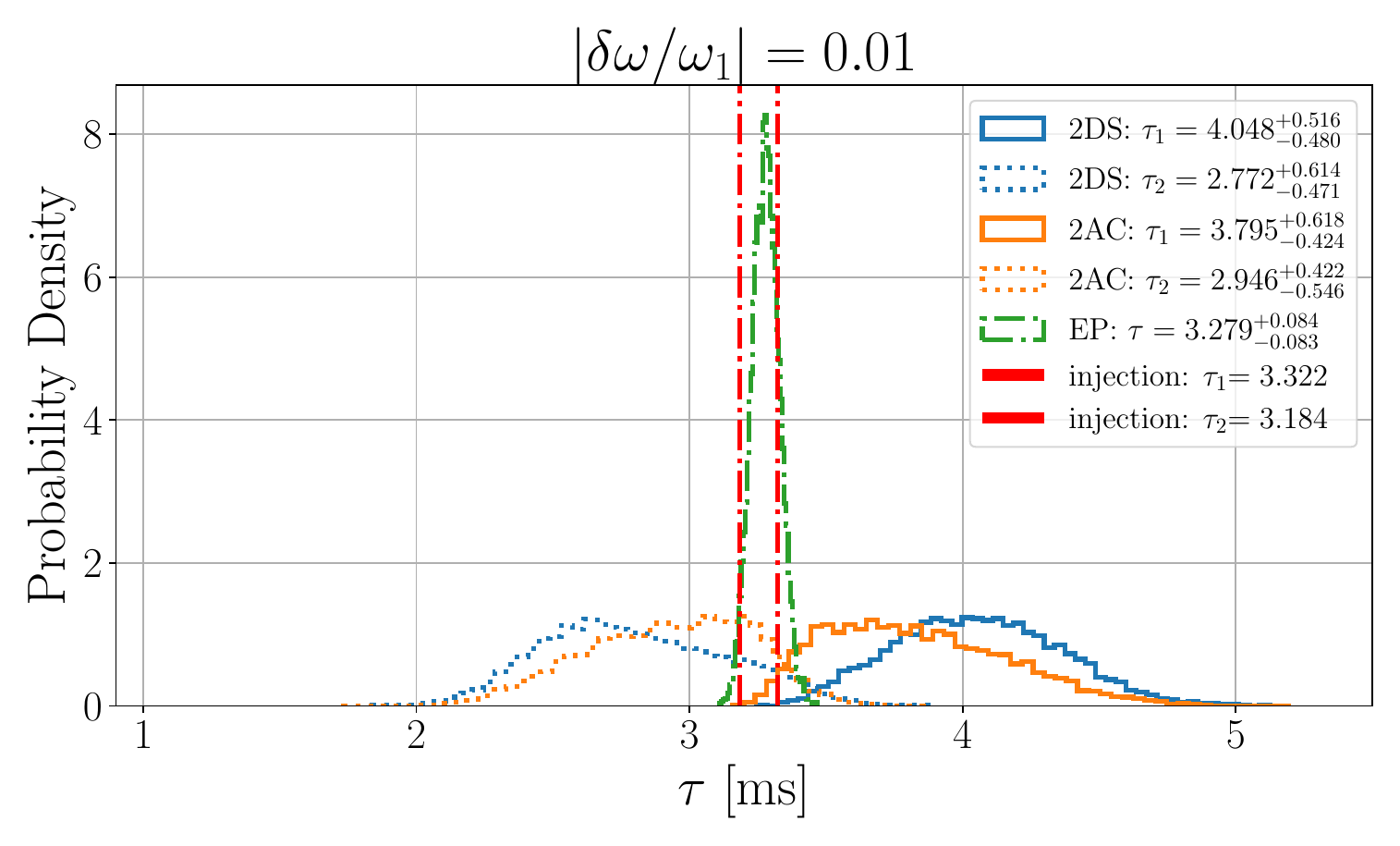}
      \end{minipage} \\
      \begin{minipage}[t]{0.45\textwidth}
        \centering
        \includegraphics[keepaspectratio, width=0.9\linewidth]{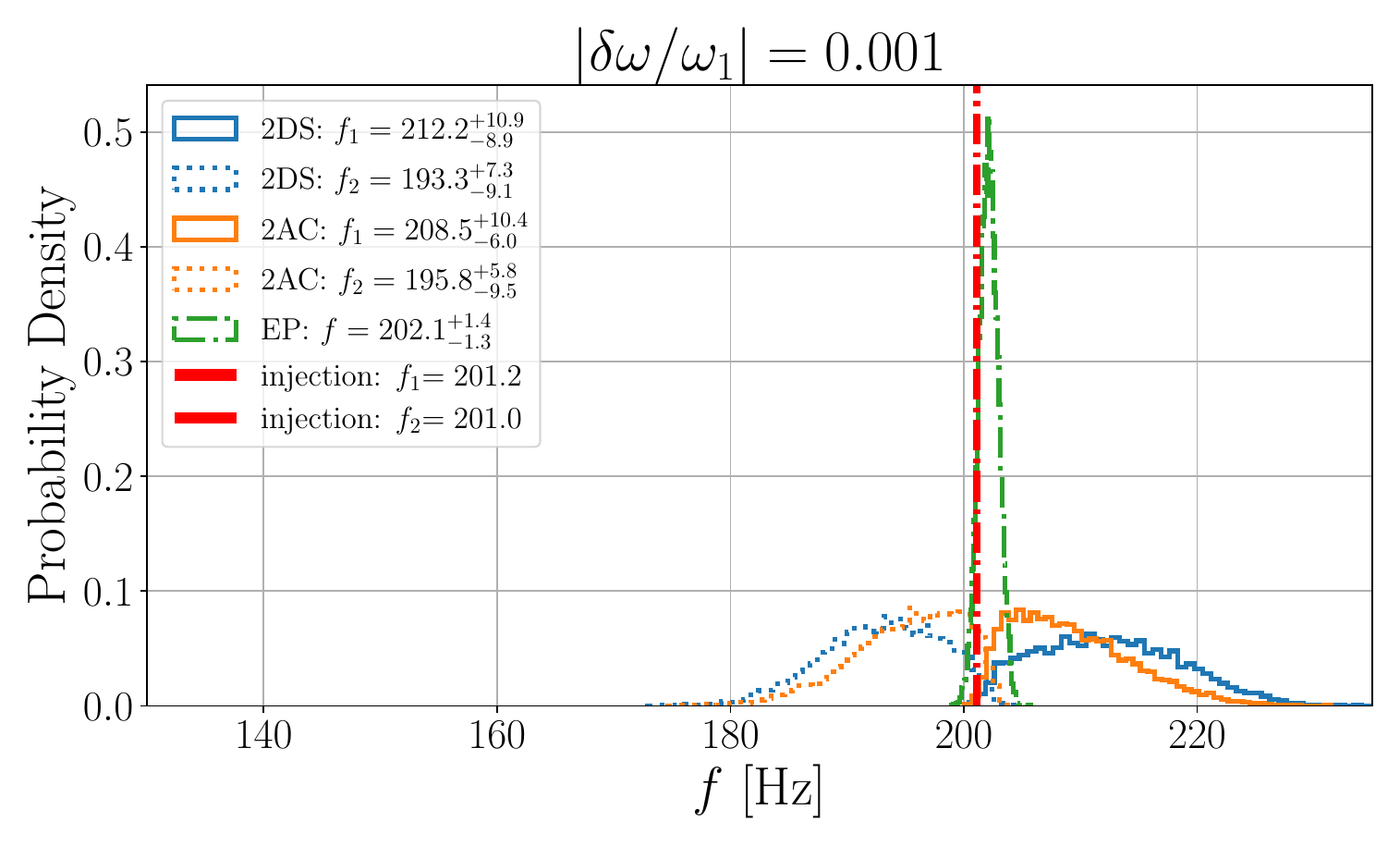}
      \end{minipage} &
      \begin{minipage}[t]{0.45\textwidth}
        \centering
        \includegraphics[keepaspectratio, width=0.9\linewidth]{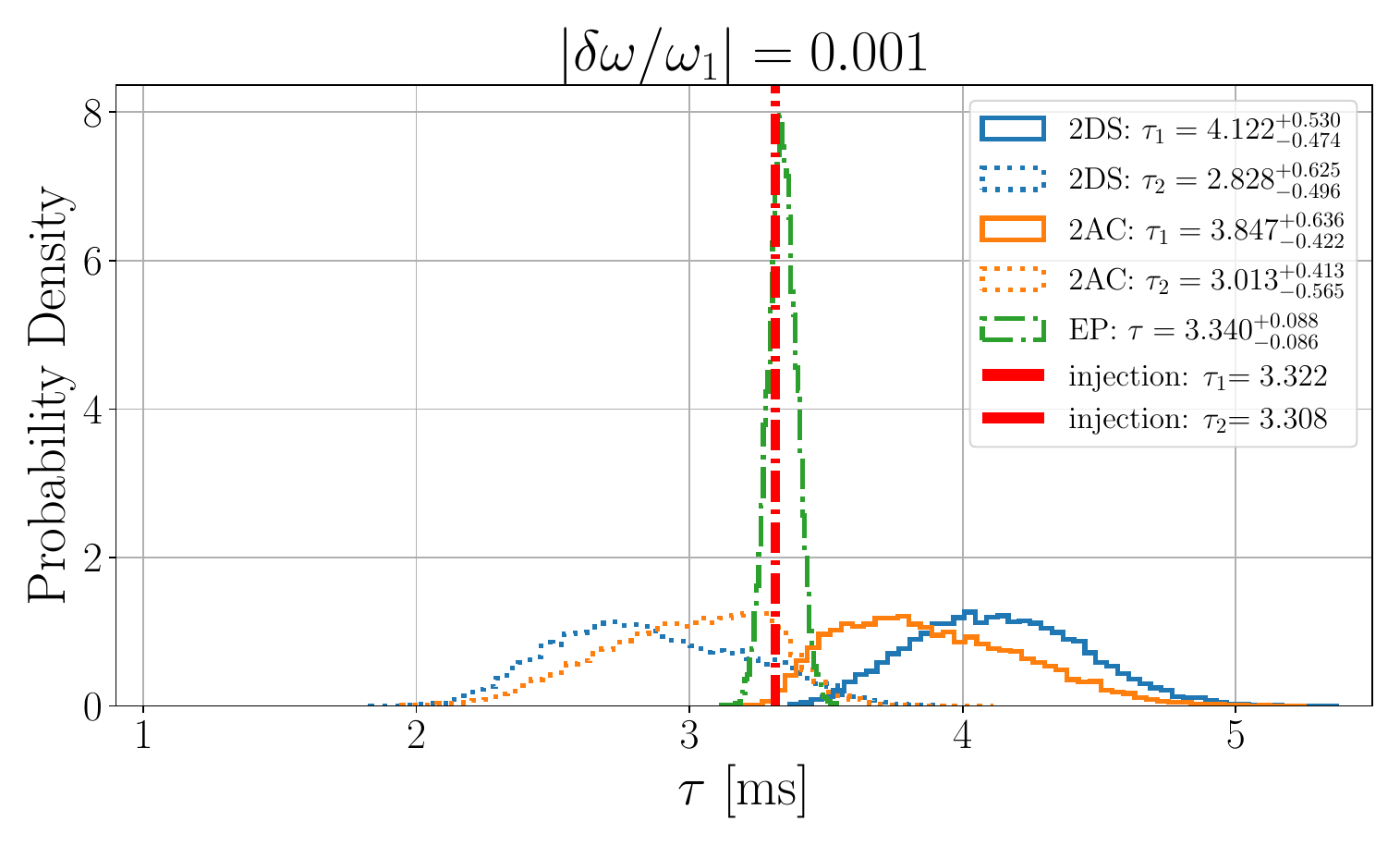}
      \end{minipage}
    \end{tabular}
    \caption{Posterior probability distributions for the frequencies and damping times. 
    The left and right columns correspond to the frequency- and damping-time-shifted cases, respectively. 
    From top to bottom, the panels correspond to the $|\delta\omega/\omega_1| = 0.3, 0.1, 0.01$, and $0.001$ cases. 
    The blue, orange, and green curves denote the posteriors obtained using the 2DS, 2AC, and EP models, respectively, while the red vertical lines indicate the injected values.
    For the 2DS and 2AC models, the solid and dashed curves correspond to the first and second modes, respectively. 
    The legends show the posterior medians and the $90\%$ credible intervals.}
    \label{fig:posterior of f and tau}
\end{figure*}

Figure~\ref{fig:posterior of f and tau} shows the posterior probability distributions of the inferred frequencies and damping times for the frequency- and damping-time-shifted injections obtained using the three waveform models. 
For $|\delta\omega/\omega_1|=0.3$, the posteriors inferred using the 2DS and 2AC models are broadly similar. 
As the complex-frequency separation $|\delta\omega/\omega_1|$ decreases, the 2AC model yields posterior peaks slightly closer to the true values, although the overall posterior widths remain comparable to those of the 2DS model.
The EP model, characterized by a single frequency and damping time, approximately recovers the injected complex frequency as an effective average of the two modes for small separations, while it fails to provide an accurate description for large separations.
This behavior is expected because the EP model is derived as an approximation valid for sufficiently small complex-frequency separations, treating the two separated QNMs as a single double-pole QNM and thereby reducing the number of parameters.


\begin{figure*}[t]
    \begin{tabular}{cc}
      \begin{minipage}[t]{0.45\textwidth}
        \centering
        \includegraphics[keepaspectratio, width=0.9\linewidth]{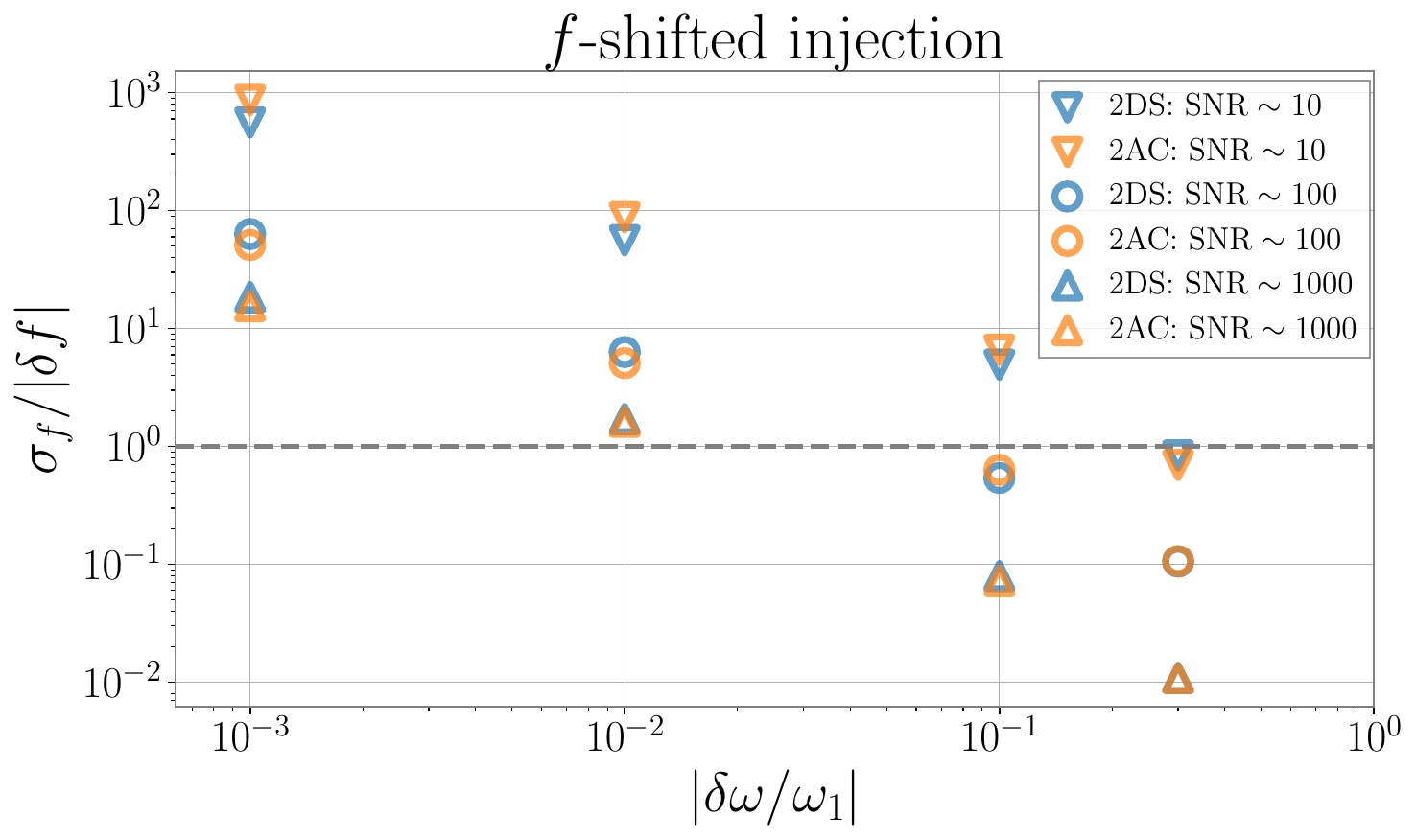}
      \end{minipage} &
      \begin{minipage}[t]{0.45\textwidth}
        \centering
        \includegraphics[keepaspectratio, width=0.9\linewidth]{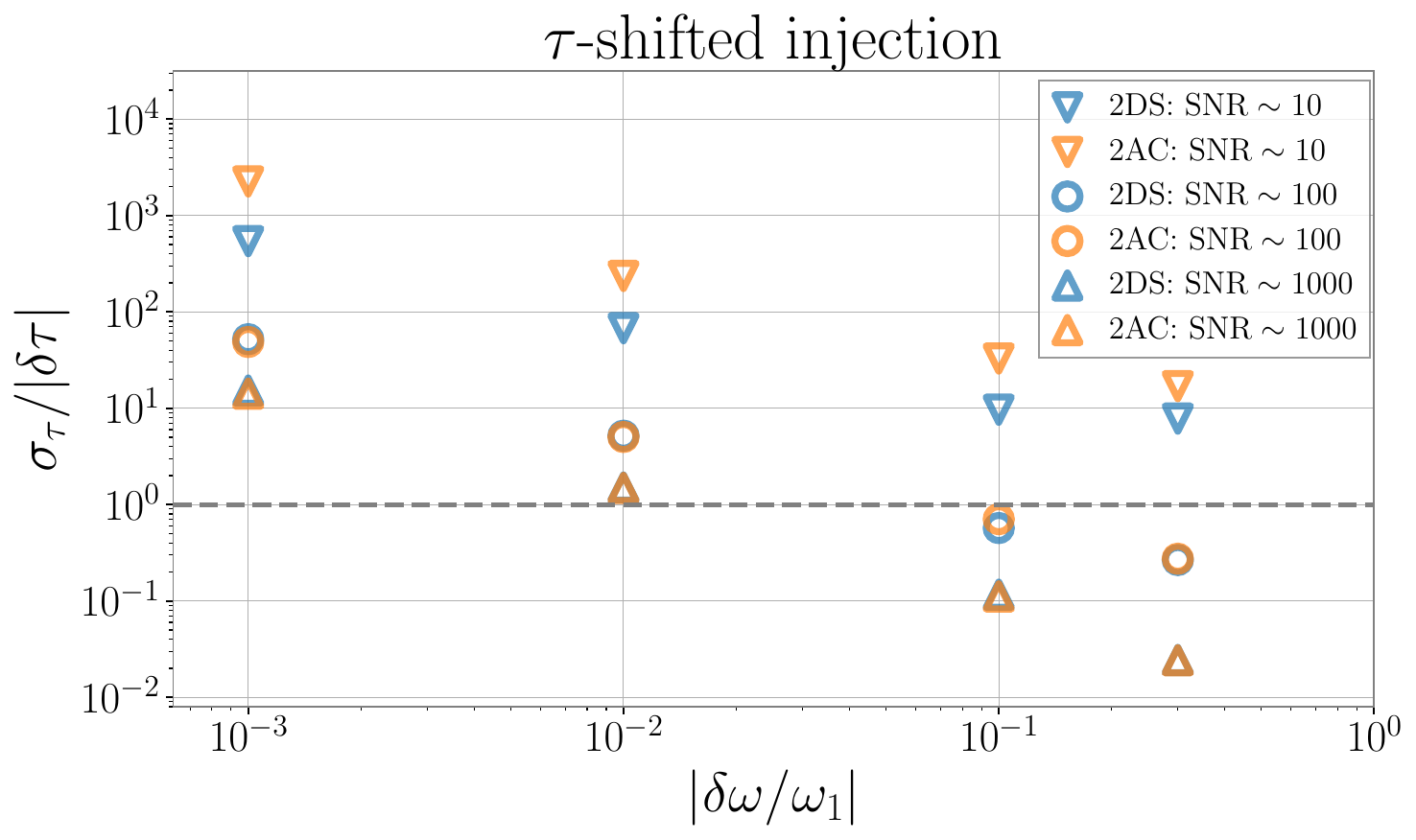}
      \end{minipage}
    \end{tabular}
    \caption{Resolvability of the frequencies (left) and damping times (right) as functions of the fractional complex-frequency separation $|\delta\omega/\omega_1|$ for the frequency- and damping-time-shifted injections, respectively. 
    The blue and orange markers correspond to the 2DS and 2AC models, respectively. 
    The marker shapes indicate the SNR: inverted triangles for $\mathrm{SNR}\sim10$, circles for $\sim 100$, and triangles for $\sim 1000$.
    The dashed horizontal line corresponds to the criterion $\sigma_f/|\delta f| = 1$ or $\sigma_\tau/|\delta\tau| = 1$, below which the two modes are regarded as spectroscopically resolvable.}
    \label{fig:resolvability for f and tau}
\end{figure*}

To quantify the resolvability of the frequencies and the damping times of two modes, we employ a Rayleigh-limit-like criterion following Ref.~\cite{Berti:2005ys},
\begin{align}
    |\delta f| &> \sigma_f \:, \\
    |\delta\tau| &> \sigma_\tau \:,
\end{align}
where $\delta f=f_1-f_2$ and $\delta\tau=\tau_1-\tau_2$ denote the differences between the true frequencies and damping times, respectively.
The quantities $\sigma_f=\mathrm{max}(\sigma_{f_1},\sigma_{f_2})$ and $\sigma_\tau=\mathrm{max}(\sigma_{\tau_1},\sigma_{\tau_2})$ represent the larger value of the inferred uncertainty for the individual errors $\sigma_{f_1},\sigma_{f_2}, \sigma_{\tau_1}$, and $\sigma_{\tau_2}$. 
This criterion requires the true frequencies and damping times to be separated by more than the corresponding posterior uncertainty. 
Figure~\ref{fig:resolvability for f and tau} shows the resolvability of the frequency and damping times for the frequency- and damping-time-shifted injections, quantified by $\sigma_f/|\delta f|$ and $\sigma_{\tau}/|\delta \tau|$, respectively. 
The uncertainties are estimated using the $90\%$ credible levels.
The condition $\sigma_f/|\delta f|$ or $\sigma_{\tau}/|\delta \tau|<1$ corresponds to the resolvable region below the gray line in the figure. 
As the complex-frequency separation decreases, both $\sigma_f/|\delta f|$ and $\sigma_{\tau}/|\delta \tau|$ increase and exceed unity around $|\delta\omega/\omega_1|\le0.1$ for $\mathrm{SNR}\sim100$. 
The posterior uncertainties decrease with increasing SNR, improving the resolvability of the modes.
Consequently, modes with smaller separations become distinguishable at higher SNRs.
However, the overall resolvability remains comparable between the 2DS and 2AC models.

To understand this behavior analytically, we perform a simplified Fisher analysis~\cite{Cutler:1994ys,Poisson:1995ef} for the 2DS model. 
We assume stationary white noise and fix the total SNR to be $\rho$.
For the frequency-shifted case, we set $\tau_1=\tau_2$ and expand the Fisher matrix in the frequency separation $\delta f$.
For the damping-time-shifted case, we set $f_1=f_2$ and expand it in the damping-time separation $\delta \tau$.
In the AC-motivated nearly degenerate regime, the two modes have comparable amplitudes and nearly opposite phases, $A_1\sim A_2$ and $\phi_1\sim\phi_2+\pi$.
Under these assumptions, the leading scaling of the resolvability ratios is
\begin{align}
\frac{\sigma_{f_i}}{|\delta f|} &\propto \frac{1}{\rho \pi^2 \tau_{i}^{2} |\delta f|^{2}} \:, \\
\frac{\sigma_{\tau_i}}{|\delta\tau|} &\propto \frac{\tau_{i}^{2}}{\rho |\delta\tau|^2} \:.
\end{align}
These estimates show that, in the nearly degenerate regime, the resolvability ratios deteriorate quadratically as the separation between the two modes decreases and improve inversely with the SNR.
This behavior accounts for the rapid loss of resolvability observed in Fig.~\ref{fig:resolvability for f and tau} at small complex-frequency separations.

\begin{figure*}[t]
    \begin{tabular}{cc}
      \begin{minipage}[t]{0.45\textwidth}
        \centering
        \includegraphics[keepaspectratio, width=0.9\linewidth]{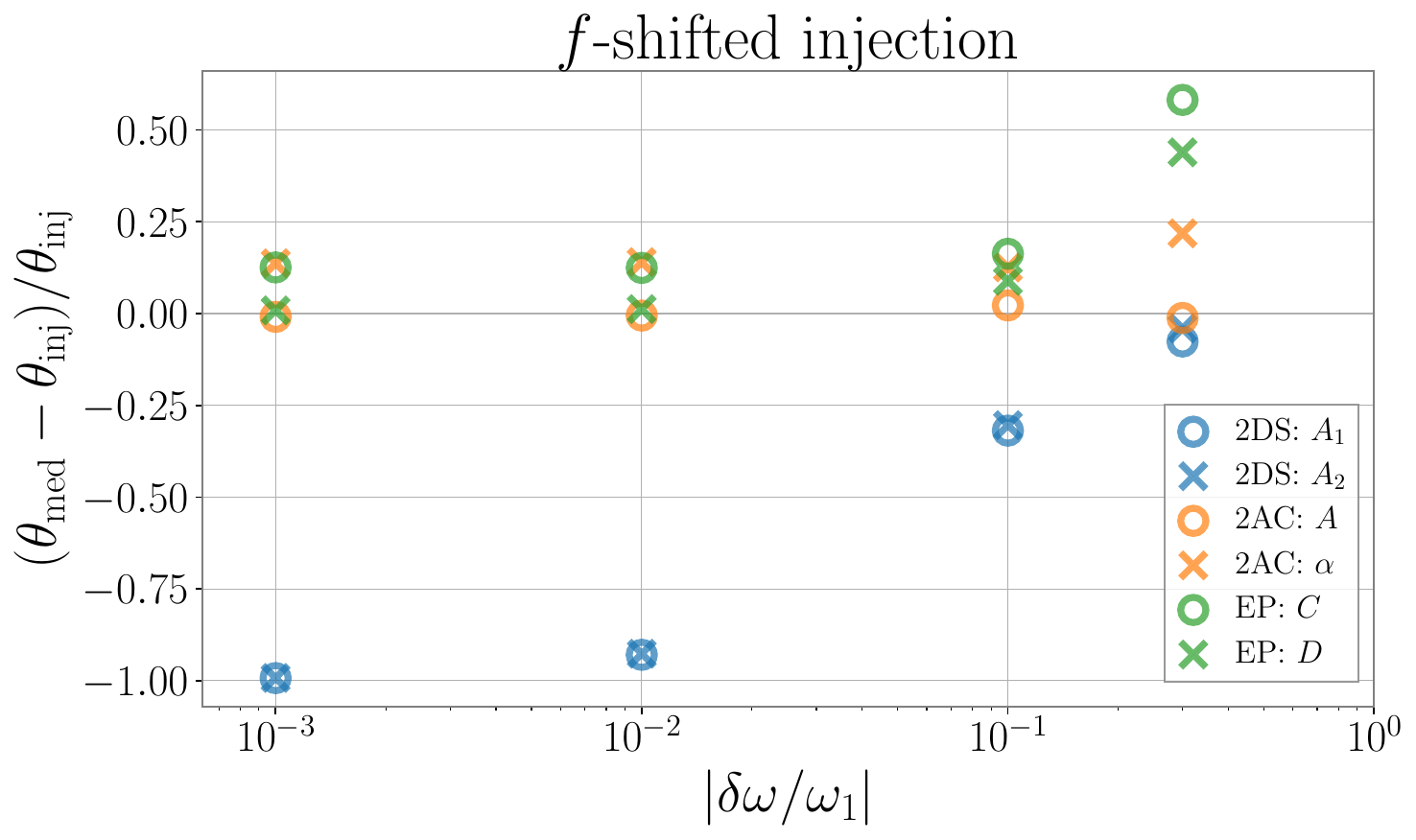}
      \end{minipage} &
      \begin{minipage}[t]{0.45\textwidth}
        \centering
        \includegraphics[keepaspectratio, width=0.9\linewidth]{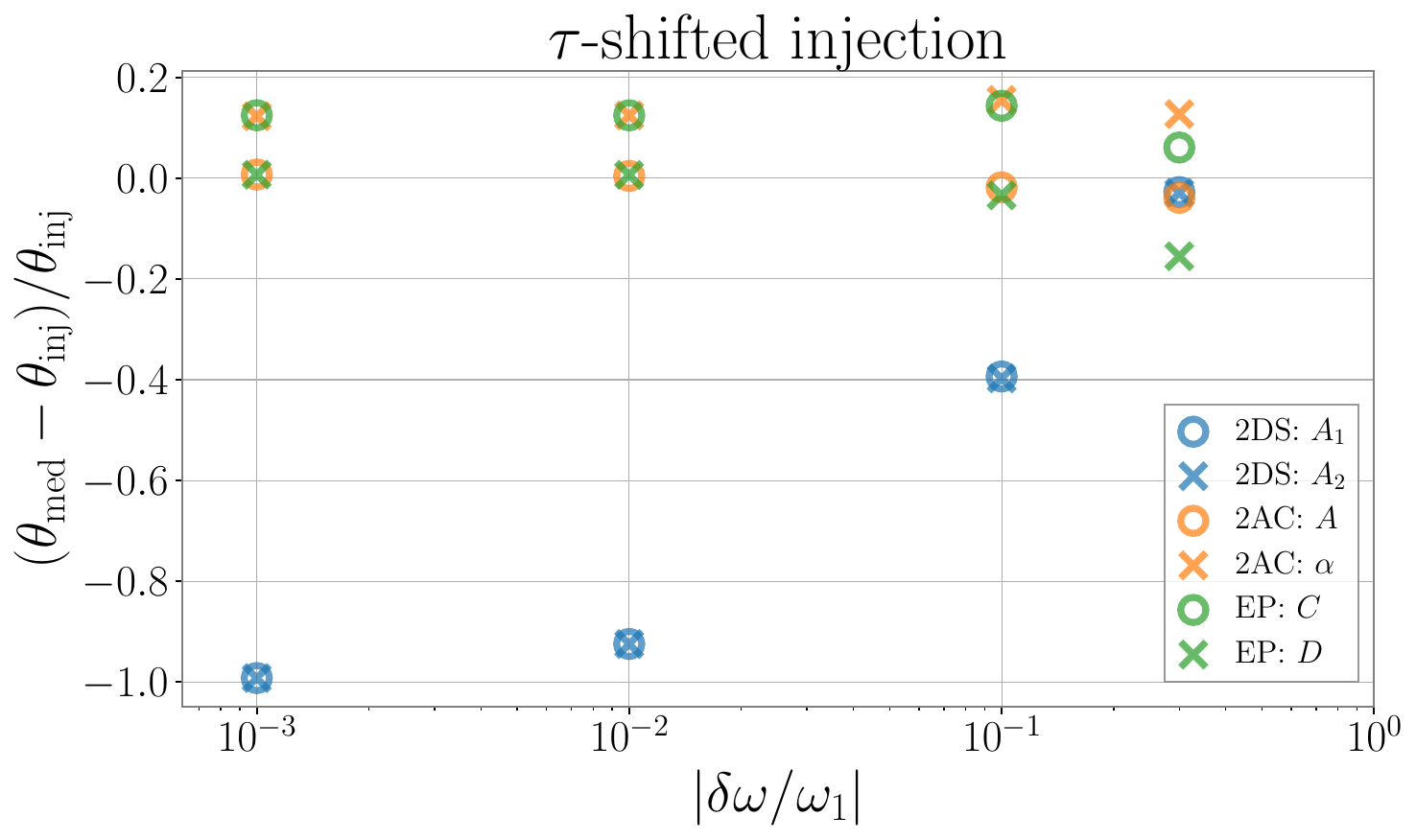}
      \end{minipage}
    \end{tabular}
    \caption{Fractional deviations of the inferred amplitude parameters from their injected values for each waveform model. 
    The left and right panels correspond to the frequency- and the damping-time-shifted injections, respectively. 
    The vertical axis shows $(\theta_{\mathrm{med}}-\theta_{\mathrm{inj}})/\theta_{\mathrm{inj}}$, where $\theta_{\mathrm{med}}$ and $\theta_{\mathrm{inj}}$ denote the median and injected values of each parameter, respectively. 
    The colors and marker styles distinguish the waveform models and amplitude parameters.}
    \label{fig:fractional accuracy for amplitudes}
\end{figure*}

The amplitude parameters are introduced differently in each waveform model. 
The 2DS model directly parametrizes the amplitudes of the individual modes, while the 2AC model incorporates the enhancement of the excitation coefficients near ACs. 
The EP model, on the other hand, parametrized the effective linear-growth behavior in the limit to the EP. 
Figure~\ref{fig:fractional accuracy for amplitudes} shows the fractional deviations of the amplitude parameters inferred using each waveform model. 
In the 2DS model, the amplitudes $A_1$ and $A_2$ are systematically underestimated for small complex-frequency separations. 
For $|\delta\omega/\omega_1| \ll 1$, the 2DS model exhibits an approximate degeneracy characterized by $A_1+A_2 \sim \mathrm{const}$. 
As a result, the signal can be reproduced by both large and small amplitudes through destructive interference between the two modes. 
However, large-amplitude solutions require finely tuned frequencies and damping times and therefore occupy only a small region in parameter space. Consequently, Bayesian inference statistically favors smaller amplitudes with broader likelihood support.

By contrast, the amplitude parameters in the 2AC and EP models are less sensitive to shifts in the other parameters. 
In the 2AC model, the parameter $A$ collectively controls the amplitudes of both modes, while in the EP model the amplitude coefficients $C$ and $D$ multiply the explicitly independent basis functions $\e^{\aye 2\pi ft}$ and $t\e^{\aye 2\pi ft}$. 
Because these parametrizations do not rely on delicate cancellations between nearly degenerate modes, the likelihood support remains broad even for large amplitudes. 
As a result, the amplitudes are inferred more accurately for small complex-frequency separations. 
The EP model provides a good approximation for sufficiently small separations, but fails to provide an accurate description for large separations, as expected from the breakdown of the double-pole approximation. 
We also confirm that the posterior distribution of $D$ remains consistent with zero when the injection consists of a single damped sinusoid without linear growth.
This supports the interpretation that the coefficient $D$ captures the EP-like linear-growth signature associated with ACs.

These results show that the detectability of AC signatures should not be identified solely with the resolvability of individual QNM frequencies.
In contrast, the amplitude-sector inference is strongly model dependent.
While the 2DS and 2AC descriptions are related by a reparametrization and yield comparable frequency resolvability, the AC-inspired parametrization provides a better-conditioned description of the collective amplitude structure.
The EP model offers a complementary diagnostic: a nonzero linear-growth coefficient indicates an effective double-pole-like waveform feature, although its interpretation as an AC signature is limited by the validity of the small-separation approximation.

\subsubsection{Application to avoided crossing in GR}
\label{subsubsec:Application to avoided crossing in GR}
\begin{figure*}[t]
    \begin{tabular}{cc}
      \begin{minipage}[t]{0.45\textwidth}
        \centering
        \includegraphics[keepaspectratio, width=0.9\linewidth]{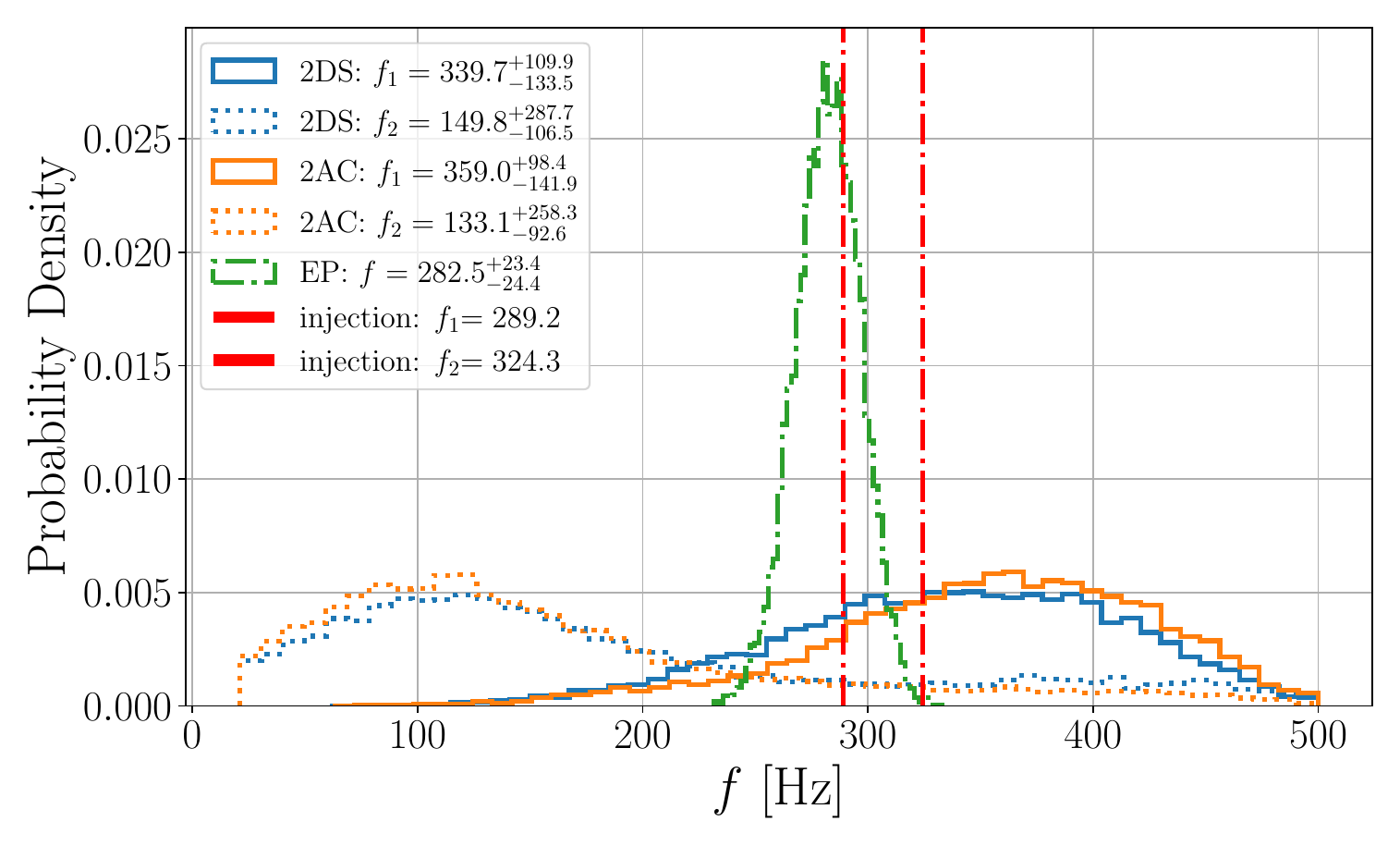}
      \end{minipage} &
      \begin{minipage}[t]{0.45\textwidth}
        \centering
        \includegraphics[keepaspectratio, width=0.9\linewidth]{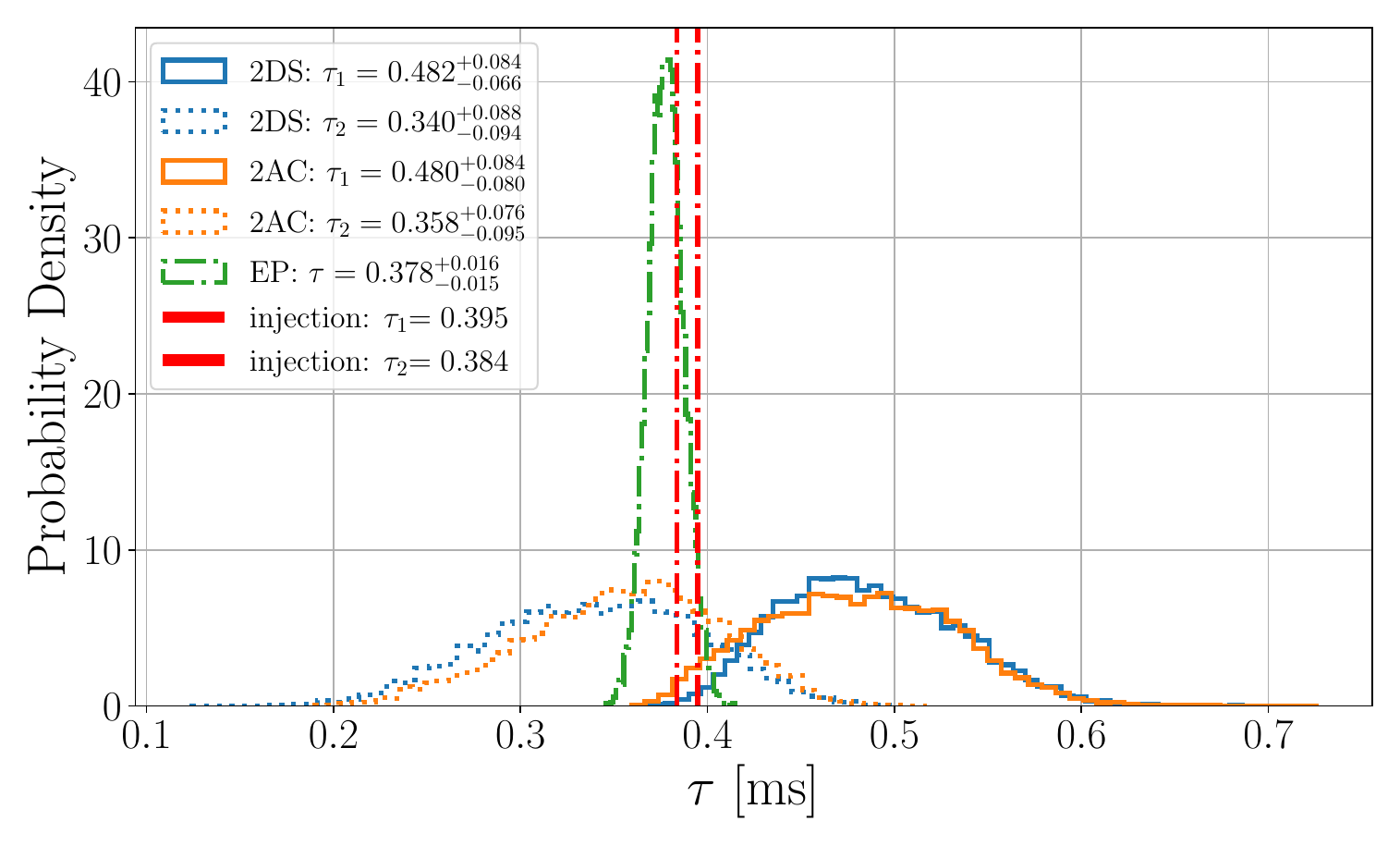}
      \end{minipage}
    \end{tabular}
    \begin{tabular}{cc}
      \begin{minipage}[t]{0.3\textwidth}
        \centering
        \includegraphics[keepaspectratio, width=1\linewidth]{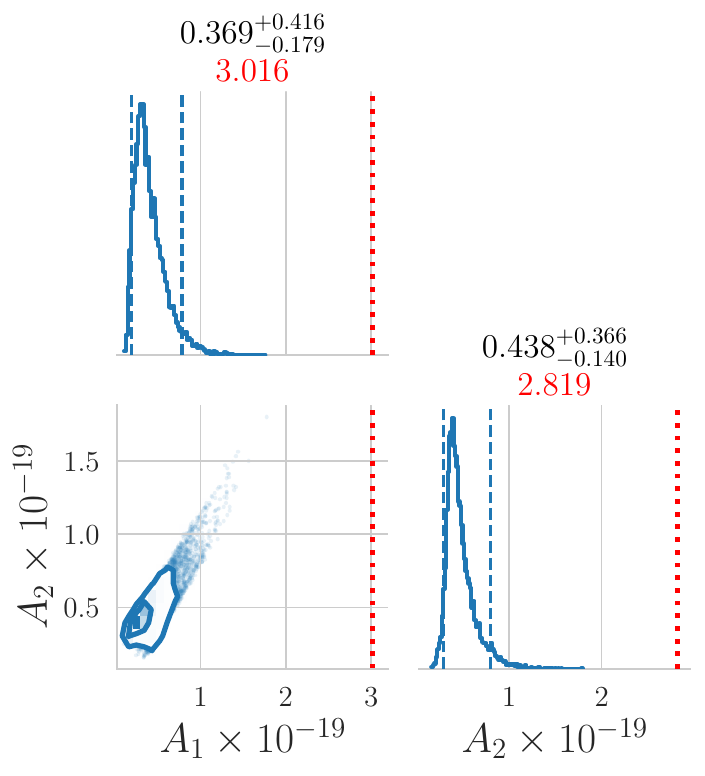}
      \end{minipage} &
      \begin{minipage}[t]{0.3\textwidth}
        \centering
        \includegraphics[keepaspectratio, width=1\linewidth]{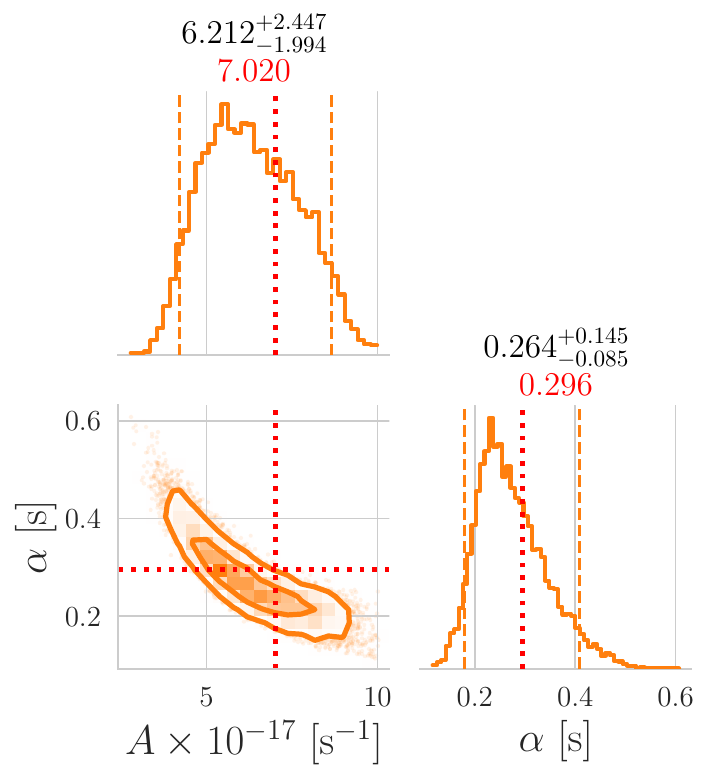}
      \end{minipage}
      \begin{minipage}[t]{0.3\textwidth}
        \centering
        \includegraphics[keepaspectratio, width=1\linewidth]{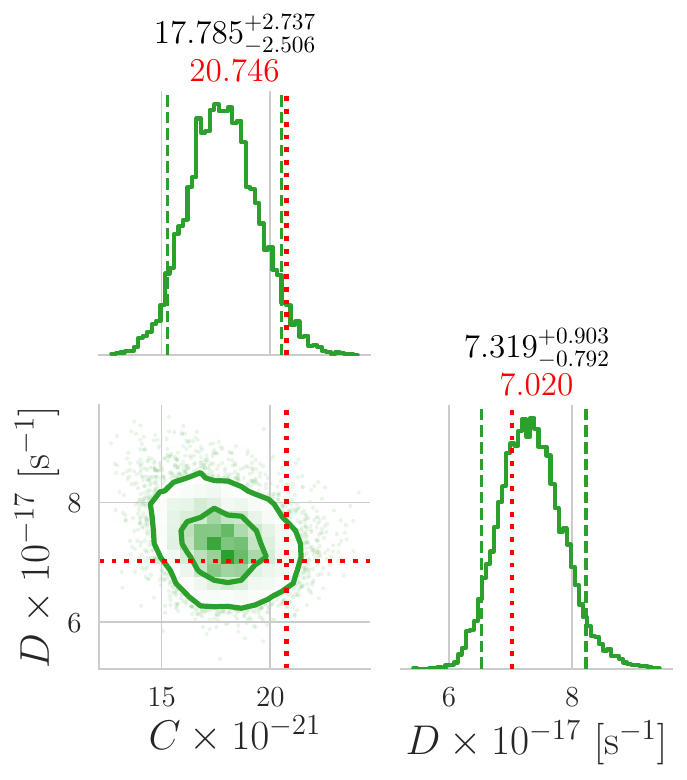}
      \end{minipage}
    \end{tabular}
    \caption{Top: the posterior distributions of the frequencies (left) and damping times (right) inferred using the 2DS (blue), 2AC (orange), and EP (green) models.
    Bottom: the corner plots of the amplitude parameters inferred using the 2DS (left), 2AC (center), and EP (right) models. The contours indicate $50\%$ and $90\%$ credible regions. 
    The red lines represent the injected values for the 2DS and 2AC models, and the corresponding effective values for the EP model derived from the relation between the AC and EP models described in Sec.~\ref{subsubsec:EP model}. 
    The black labels indicate the posterior medians and the $90\%$ credible intervals.}
    \label{fig:posteriors for 225 and 226}
\end{figure*}

As an example of the AC in GR, we apply the three waveform models to an idealized ringdown signal composed of the $(\ell,m,n)=(2,2,5)$ and $(2,2,6)$ QNMs of a Kerr black hole with dimensionless spin $j=0.9$ and mass $M=60~M_{\odot}$. 
Here we inject these two modes and assume that the lower overtones have been removed beforehand.
This idealized setup allows us to assess the detectability of AC signatures under optimistic conditions.

The injected frequencies and damping times are set to $f_1\sim289.2$ and $f_2\sim324.3~\mathrm{Hz}$, $\tau_1\sim0.394$, and $\tau_2\sim0.383~\mathrm{ms}$, which yield $|\delta\omega/\omega_1|\sim0.075$. 
Accordingly, the lower bound of the prior for the damping time is reduced to $5\times10^{-5}~\mathrm{s}$ in this analysis.
As in the controlled injections, the signal is generated using the 2AC model.
The amplitude is chosen such that the two-mode signal composed only of the (2,2,5) and (2,2,6) modes has $\mathrm{SNR}\sim100$. 
We set $\alpha=M$ with $M=60M_{\odot}$, corresponding to $\alpha \sim 0.296~\mathrm{ms}$, and choose $\phi_A=\phi_{\alpha}=0$.

Figure~\ref{fig:posteriors for 225 and 226} shows the posterior distributions inferred using the three waveform models. 
The initial phase parameters are not shown since they are not relevant to the present discussion. 
The posteriors of the frequencies and damping times obtained with the 2DS and 2AC models are qualitatively similar to those in the controlled injections.
The corresponding resolvability measures are $(\sigma_f/|\delta f|,\sigma_\tau/|\delta \tau|)\sim(7.305,10.333)$ for the 2DS model and $(5.457,9.579)$ for the 2AC model. 
Thus, even in this idealized two-mode setup with high SNR, the individual complex frequencies are not spectroscopically resolved.
The amplitude-sector inference shows the same tendency as in the controlled injections.
The EP model recovers the effective values predicted by the small-separation approximation, namely, $C=A\alpha$, $D=A$, $f=f_1$, and $\tau=\tau_1$, inferring an effective frequency and damping time associated with the nearly degenerate QNM pair.
Notably, the coefficient $D$ is inferred to be nonzero, indicating that the EP waveform captures the linear-growth component expected from the AC. 
We also verify that when the injected signal consists of a single damped sinusoid with the same SNR, the posterior of $D$ is consistent with zero.

Even for this optimistic GR example of the AC between $(2,2,5)$ and $(2,2,6)$, where lower ringdown modes are assumed to be removed beforehand, the individual complex frequencies are not spectroscopically resolved.
Nevertheless, the EP-inspired waveform model recovers a nonzero linear-growth coefficient, indicating that effective signatures of nearly degenerate mode interference can still remain observable.
In summary, our results suggest that (i) resolving the $(2,2,5)$ and $(2,2,6)$ modes is challenging even in the optimistic situation, but (ii) their interference signature (linear growth in time), unique to the AC, may be captured by the 2AC or EP model, provided that other long-lived QNMs or overtones irrelevant to the AC can be removed from a ringdown waveform.

\section{Discussion and Conclusion}
\label{sec:Discussion_Conclusion}
This paper investigates the detectability of AC signatures in black hole ringdown signals using three waveform models within a Bayesian framework.
Our analysis focused on the inference of nearly degenerate QNM pairs and on the extent to which characteristic waveform signatures remain observable when the individual complex frequencies become difficult to resolve.

For controlled injections of nearly degenerate QNM pairs, the 2AC model slightly improves the recovery of the injected parameters compared with the 2DS model.
However, the intrinsic resolvability of the individual frequencies and damping times remains limited even under optimistic conditions.
For the representative setup with $\mathrm{SNR}=100$, the transition between resolvable and unresolved regimes occurs around $|\delta\omega/\omega_1|\sim0.1$, although this threshold depends on the SNR and the damping time of the modes.
The EP model provides an effective description only within the regime where the double-pole approximation remains valid.


The inference of the amplitude parameters depends more strongly on the waveform parametrization.
In the 2DS model, the amplitudes are systematically underestimated for small complex-frequency separations because large-amplitude solutions require finely tuned cancellations between two modes and therefore occupy only a small region in parameter space. 
Consequently, Bayesian inference statistically favors smaller amplitudes with broader likelihood support.
By contrast, the 2AC and EP parametrizations provide a better-conditioned description of the collective amplitude structure and therefore recover the injected amplitudes more accurately in the nearly degenerate regime.


We also applied the waveform models to a GR example of the AC between $(2,2,5)$ and $(2,2,6)$ for a Kerr black hole with $j = 0.9$.
Even in this optimistic setup, where only two modes are injected with high SNRs and the lower overtones are assumed to be removed beforehand, the individual complex frequencies are not resolved.
Nevertheless, the EP-inspired waveform model recovers a nonzero linear-growth coefficient $D$, indicating that effective signatures of nearly degenerate mode interference can remain observationally accessible in principle.


Our analysis suggests that the detectability of AC signatures should not be identified solely with the resolvability of individual QNM frequencies.
Although the intrinsic frequency resolvability remains limited even with AC-inspired waveform models, collective waveform features associated with nearly degenerate mode interference can still remain inferable through suitable parametrization.

In particular, our 2AC model enables us to identify the key factors that determine when the linear growth arising from QNM interference becomes dominant in the presence of an AC \cite{Nakamoto:2026lyo}.
The EP-inspired waveform provides a useful effective description in the unresolved regime \cite{Yang:2025dbn,PanossoMacedo:2025xnf}, although its interpretation is restricted to the validity range of the small-separation approximation.



The present analysis was intentionally optimistic in order to clarify the fundamental detectability limits associated with nearly degenerate QNMs, employing high SNR and idealized two-mode injections.
In realistic observations, additional lower overtones, waveform systematics, and detector noise will further complicate the extraction of nearly degenerate modes~\cite{Dreyer:2003bv, Baibhav:2017jhs, Yang:2017zxs}.
Nevertheless, our results indicate that combining multiple waveform descriptions may provide a practical strategy for probing non-Hermitian physics in future black hole spectroscopy analyses.


\section*{Acknowledgments}
\label{sec:Acknowledgments}
We thank Daiki Watarai for useful discussions.
H.~I. is supported by JST SPRING, Grant No.~JPMJSP2108.
N.~O. was supported by Japan Society
for the Promotion of Science (JSPS) KAKENHI Grants
No.~JP23K13111 and No.~JP26K17142.
H.~T. is supported by the Hakubi project at Kyoto University, and by JSPS KAKENHI Grants No. JP22K14037 and No.~JP26K17146. 

\appendix
\section{Posterior distributions of amplitude parameters in controlled injections}
\label{sec:Amplitude posteriors for agnostic injection}
\begin{figure*}[t]
    \begin{tabular}{cc}
      \begin{minipage}[t]{0.28\textwidth}
        \centering
        \includegraphics[keepaspectratio, width=1\linewidth]{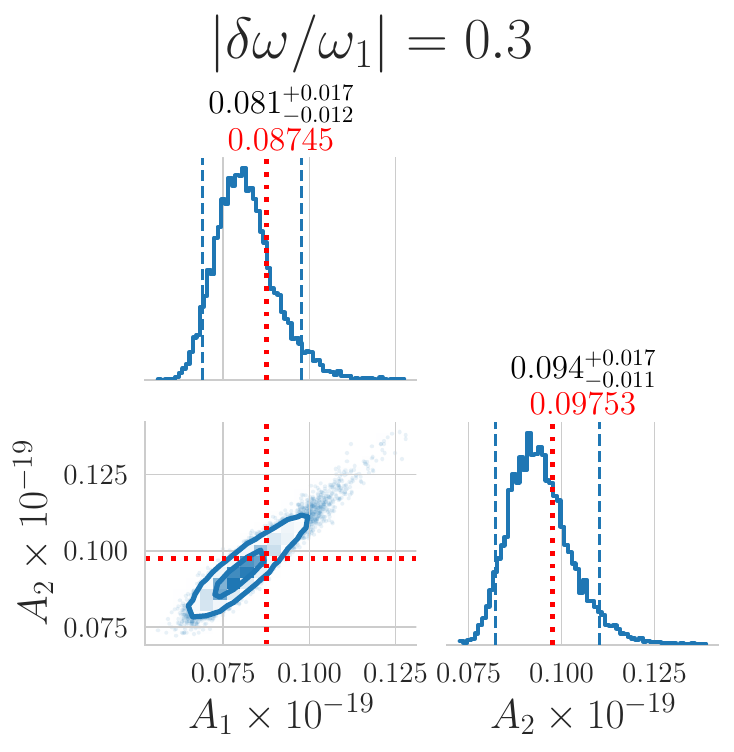}
      \end{minipage} &
      \begin{minipage}[t]{0.28\textwidth}
        \centering
        \includegraphics[keepaspectratio, width=1\linewidth]{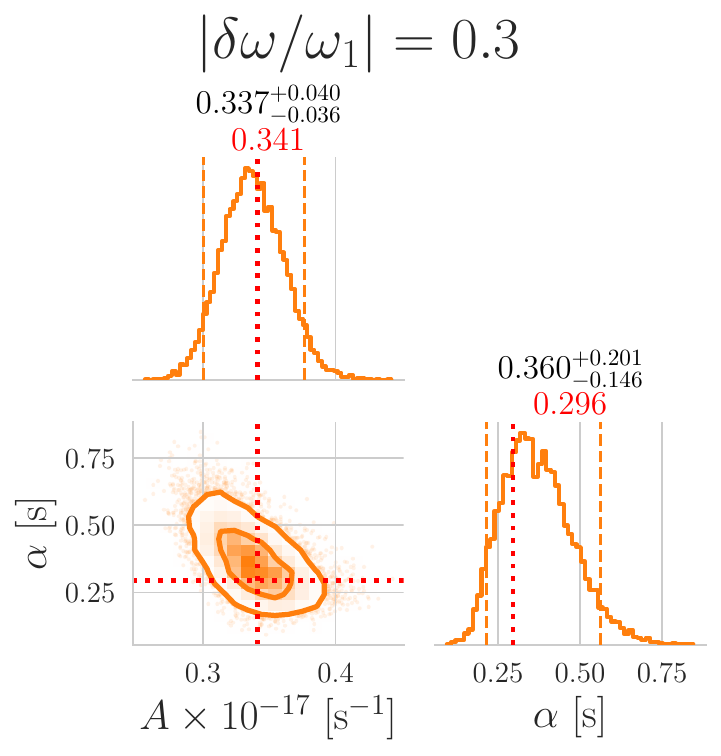}
      \end{minipage}
      \begin{minipage}[t]{0.28\textwidth}
        \centering
        \includegraphics[keepaspectratio, width=1\linewidth]{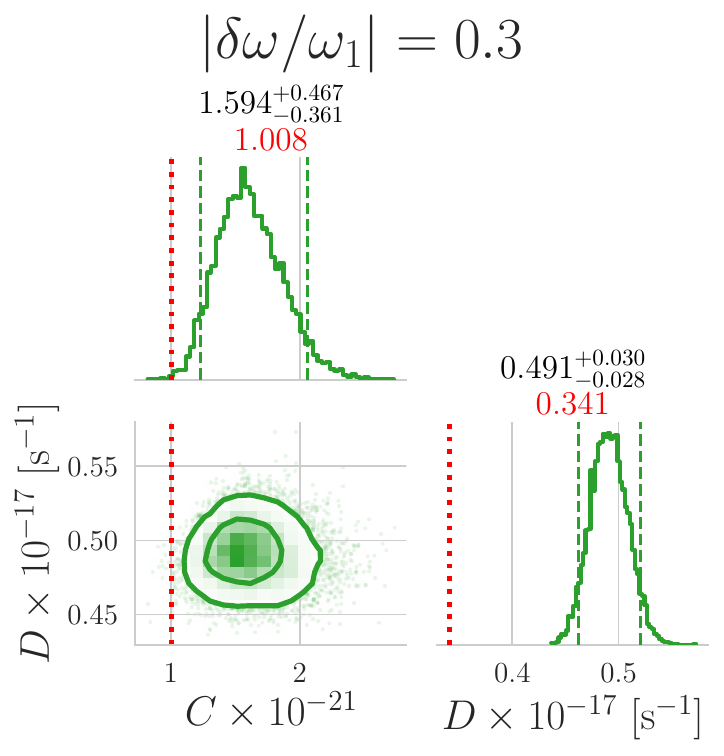}
      \end{minipage} \\
      \begin{minipage}[t]{0.28\textwidth}
        \centering
        \includegraphics[keepaspectratio, width=1\linewidth]{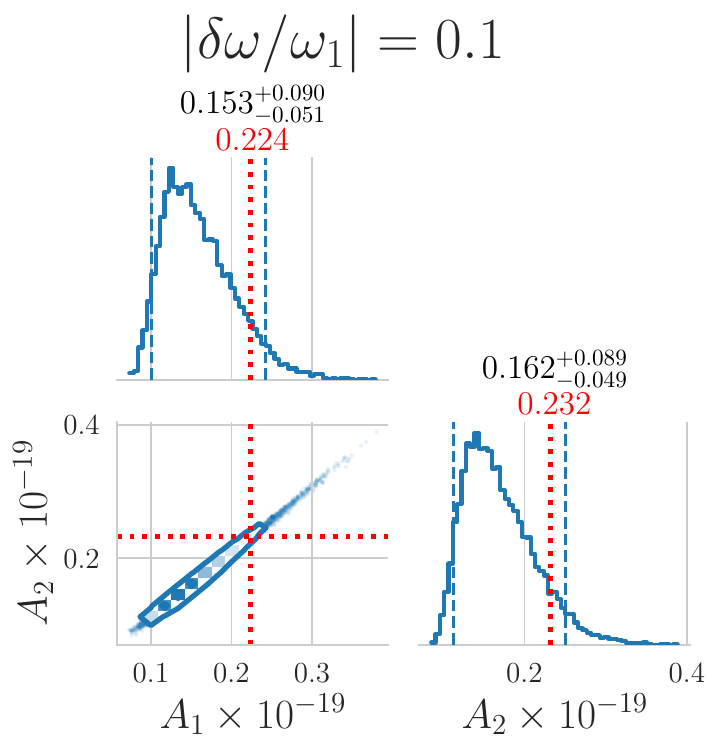}
      \end{minipage} &
      \begin{minipage}[t]{0.28\textwidth}
        \centering
        \includegraphics[keepaspectratio, width=1\linewidth]{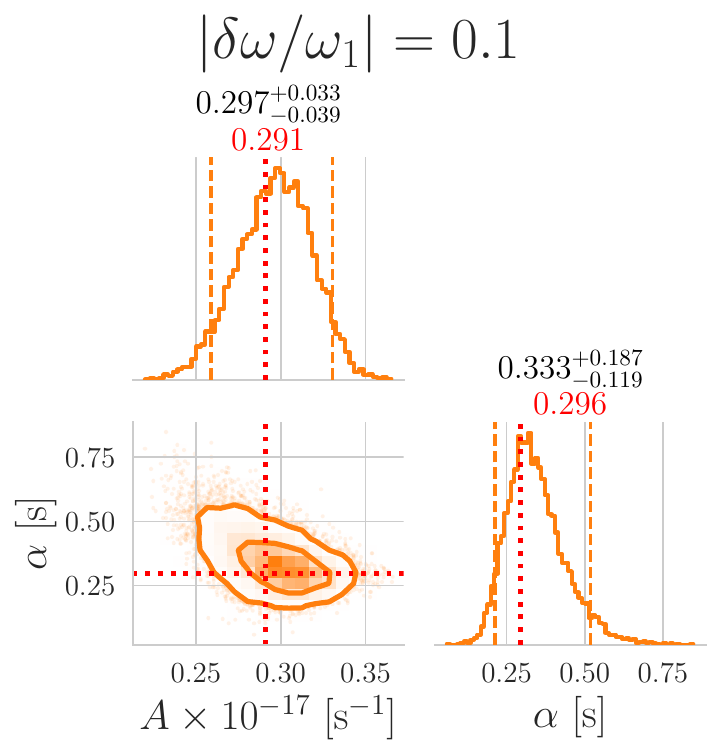}
      \end{minipage}
      \begin{minipage}[t]{0.28\textwidth}
        \centering
        \includegraphics[keepaspectratio, width=1\linewidth]{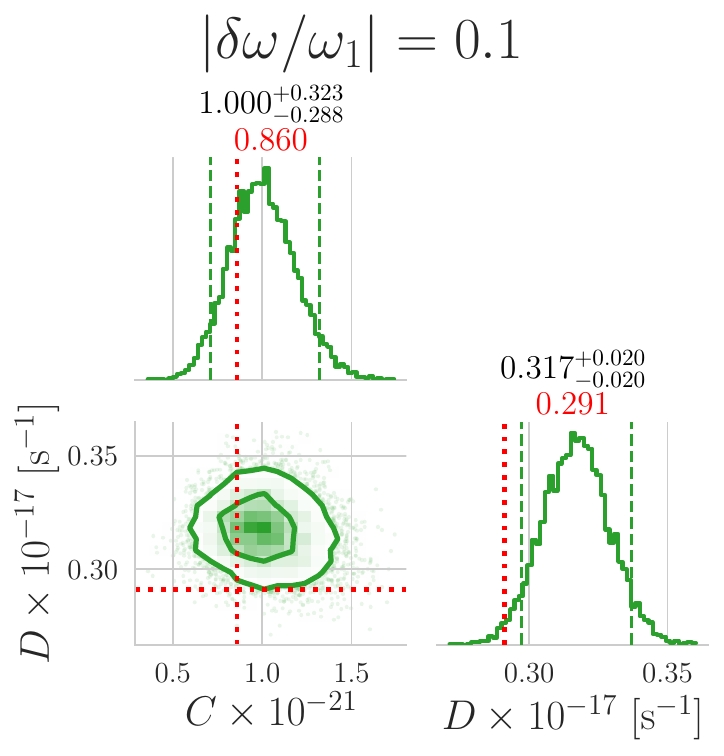}
      \end{minipage} \\
      \begin{minipage}[t]{0.28\textwidth}
        \centering
        \includegraphics[keepaspectratio, width=1\linewidth]{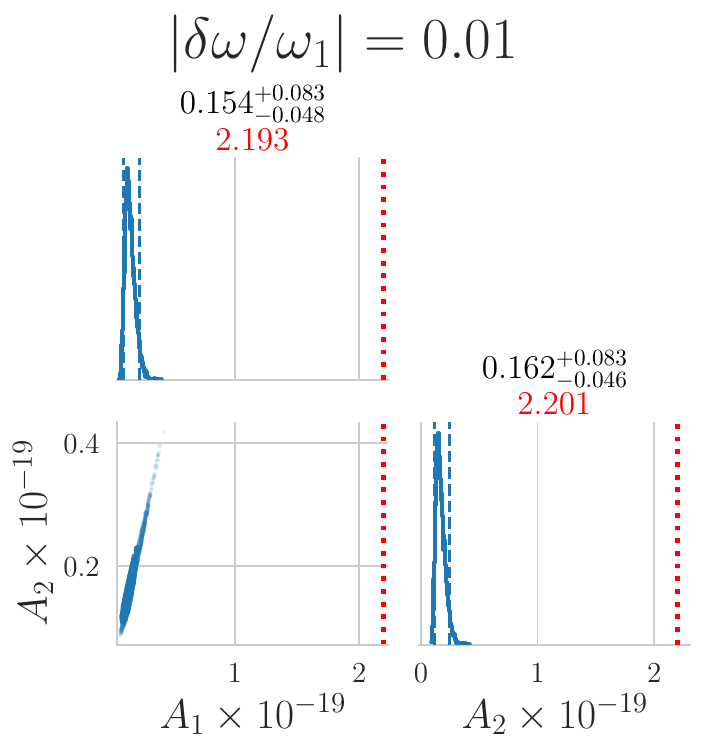}
      \end{minipage} &
      \begin{minipage}[t]{0.28\textwidth}
        \centering
        \includegraphics[keepaspectratio, width=1\linewidth]{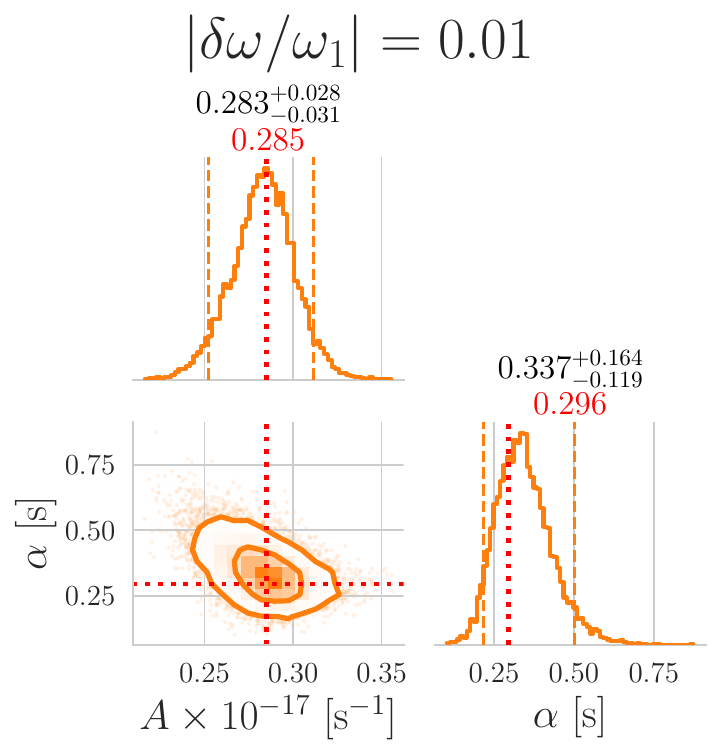}
      \end{minipage}
      \begin{minipage}[t]{0.28\textwidth}
        \centering
        \includegraphics[keepaspectratio, width=1\linewidth]{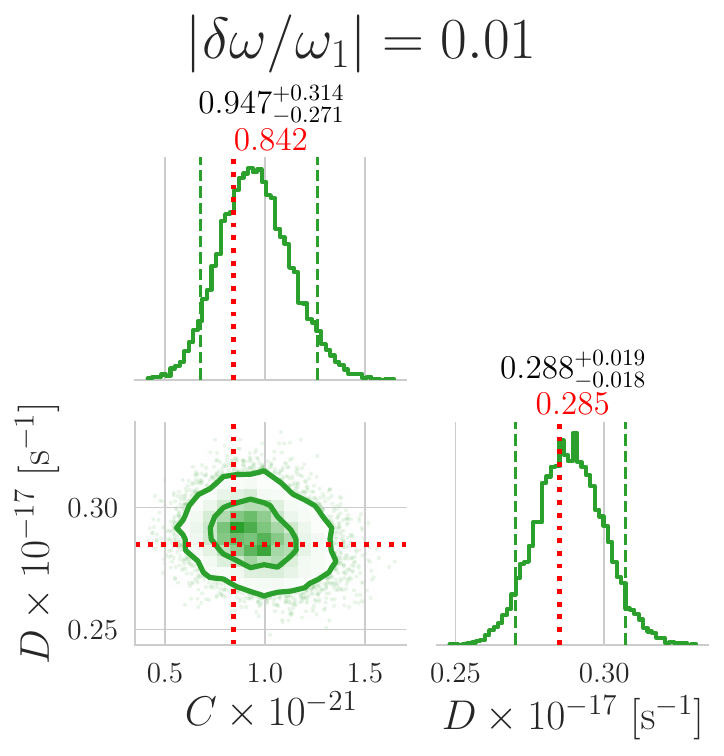}
      \end{minipage} \\
      \begin{minipage}[t]{0.28\textwidth}
        \centering
        \includegraphics[keepaspectratio, width=1\linewidth]{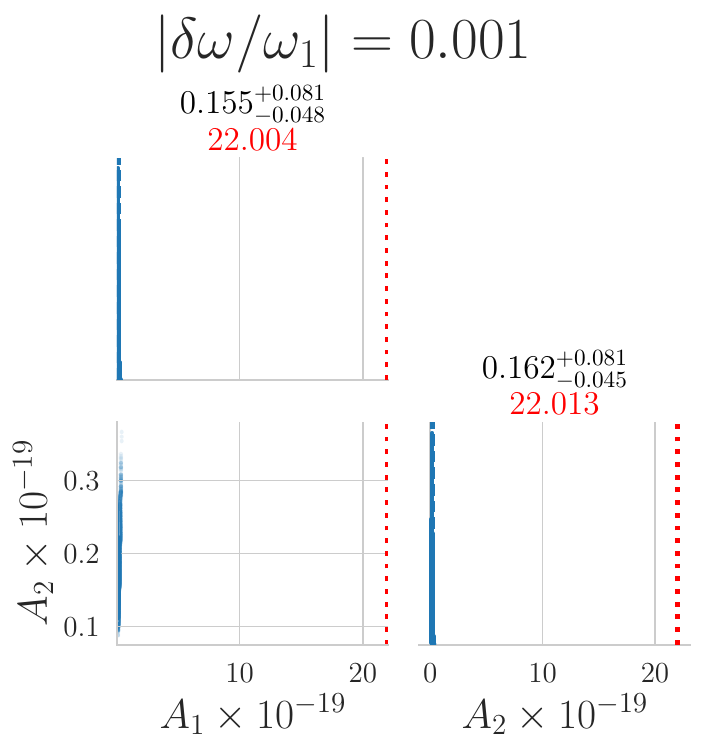}
      \end{minipage} &
      \begin{minipage}[t]{0.28\textwidth}
        \centering
        \includegraphics[keepaspectratio, width=1\linewidth]{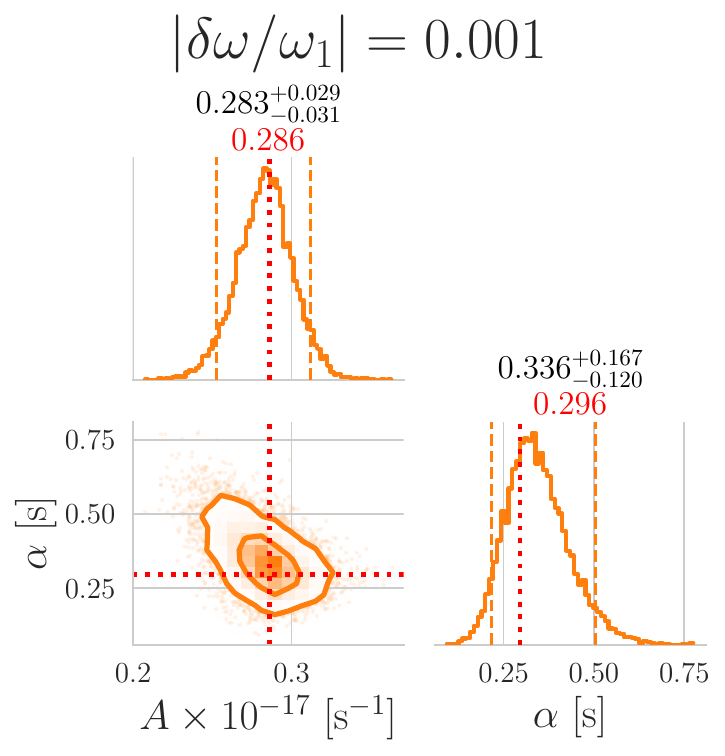}
      \end{minipage}
      \begin{minipage}[t]{0.28\textwidth}
        \centering
        \includegraphics[keepaspectratio, width=1\linewidth]{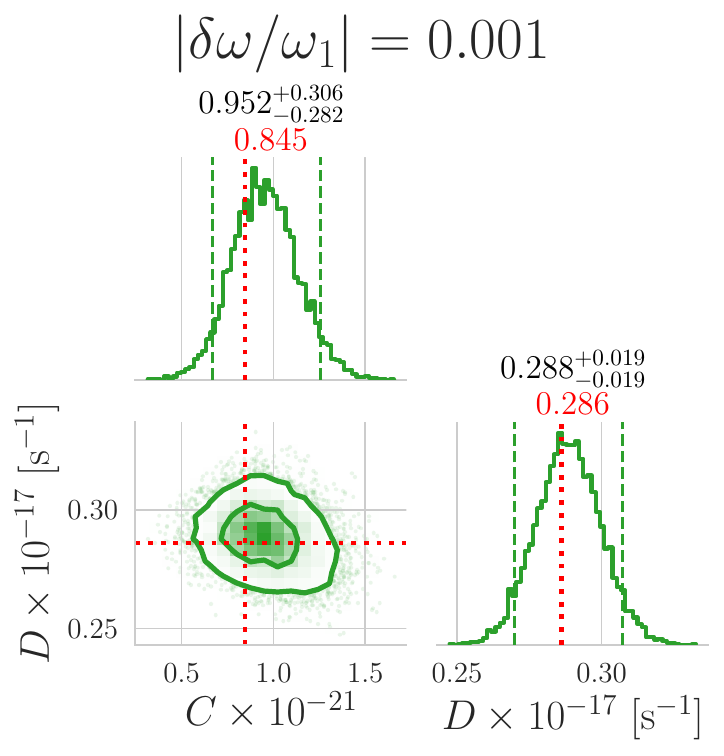}
      \end{minipage}
    \end{tabular}
    \caption{Posterior probability distributions of the amplitude parameters for the frequency-shifted injection. The left, center, and right columns correspond to the 2DS, 2AC, and EP models, respectively. 
    From top to bottom, the panels correspond to $|\delta\omega/\omega_1|=0.3$, $0.1$, $0.01$, and $0.001$.
    The red dashed lines indicate the injected values for the 2DS and 2AC models, and the corresponding effective values for the EP model derived from the AC-EP correspondence described in Sec.~\ref{subsubsec:EP model}.
    The black labels indicate the posterior medians and the $90\%$ credible intervals.
}
    \label{fig:corner plot for amplitudes in frequency shifted injection}
\end{figure*}

\begin{figure*}[t]
    \begin{tabular}{cc}
      \begin{minipage}[t]{0.28\textwidth}
        \centering
        \includegraphics[keepaspectratio, width=1\linewidth]{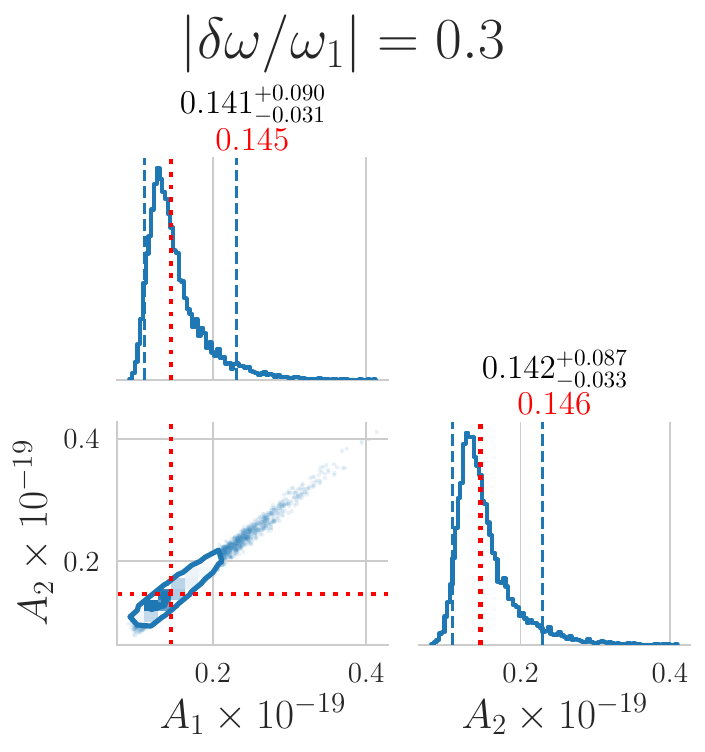}
      \end{minipage} &
      \begin{minipage}[t]{0.28\textwidth}
        \centering
        \includegraphics[keepaspectratio, width=1\linewidth]{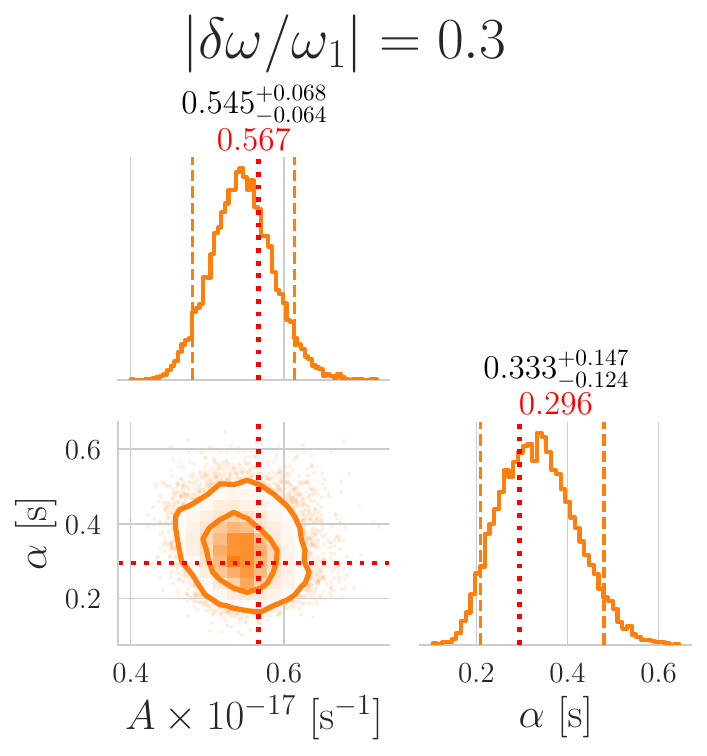}
      \end{minipage}
      \begin{minipage}[t]{0.28\textwidth}
        \centering
        \includegraphics[keepaspectratio, width=1\linewidth]{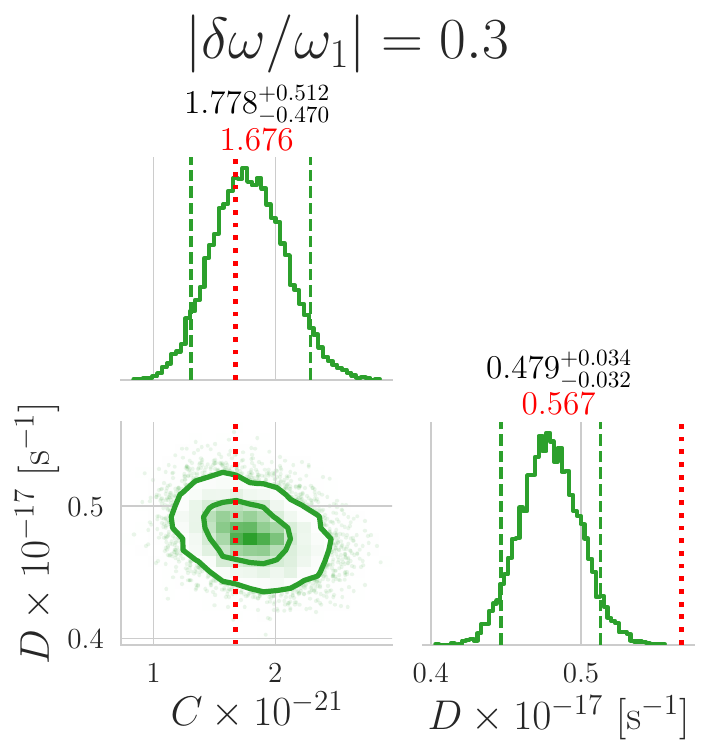}
      \end{minipage} \\
      \begin{minipage}[t]{0.28\textwidth}
        \centering
        \includegraphics[keepaspectratio, width=1\linewidth]{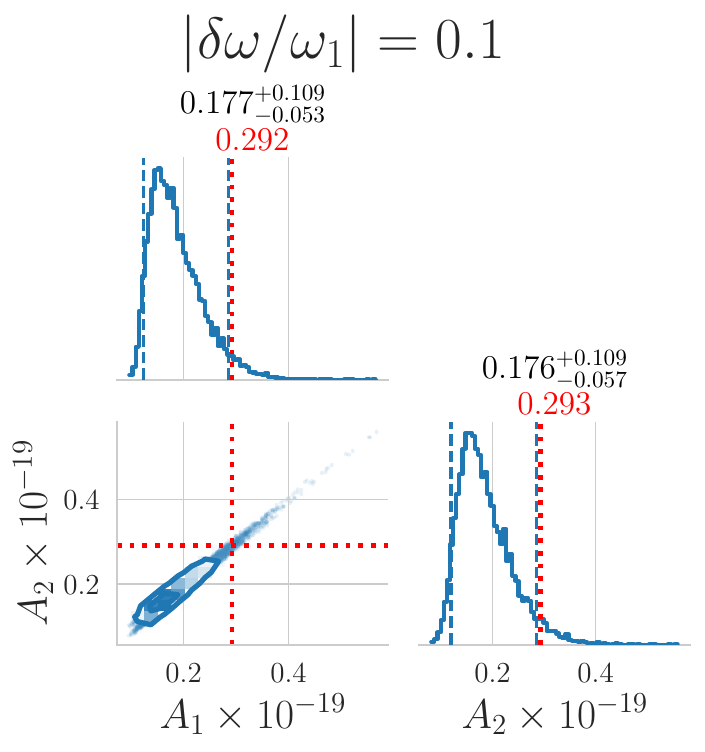}
      \end{minipage} &
      \begin{minipage}[t]{0.28\textwidth}
        \centering
        \includegraphics[keepaspectratio, width=1\linewidth]{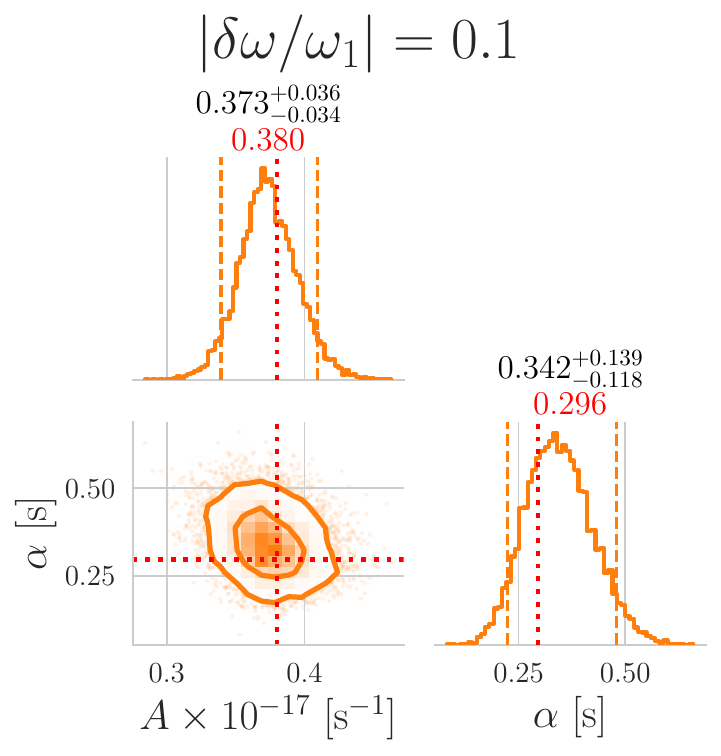}
      \end{minipage}
      \begin{minipage}[t]{0.28\textwidth}
        \centering
        \includegraphics[keepaspectratio, width=1\linewidth]{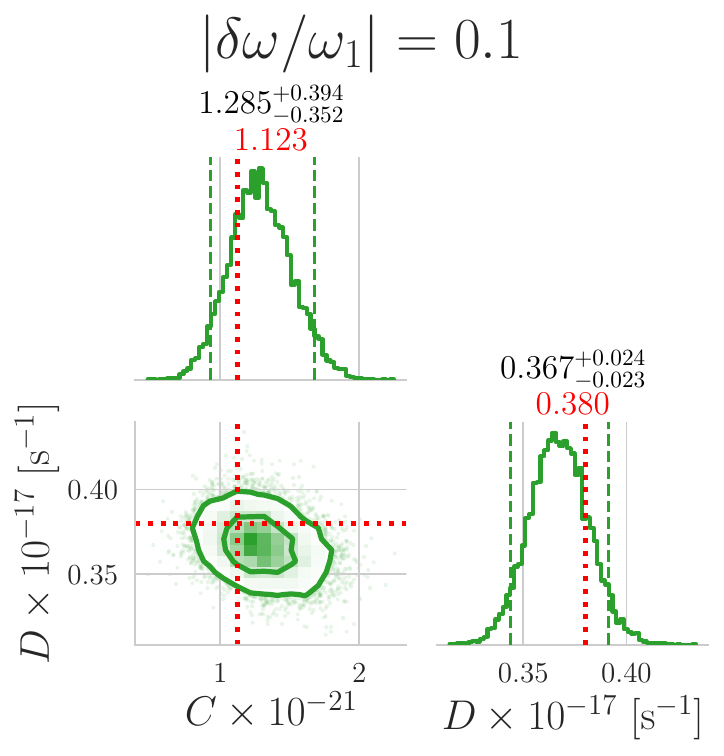}
      \end{minipage} \\
      \begin{minipage}[t]{0.28\textwidth}
        \centering
        \includegraphics[keepaspectratio, width=1\linewidth]{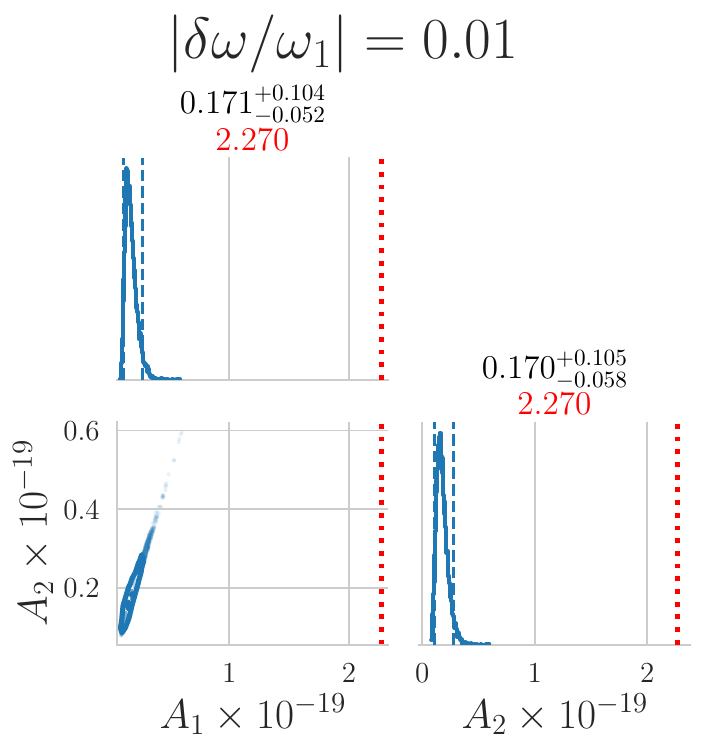}
      \end{minipage} &
      \begin{minipage}[t]{0.28\textwidth}
        \centering
        \includegraphics[keepaspectratio, width=1\linewidth]{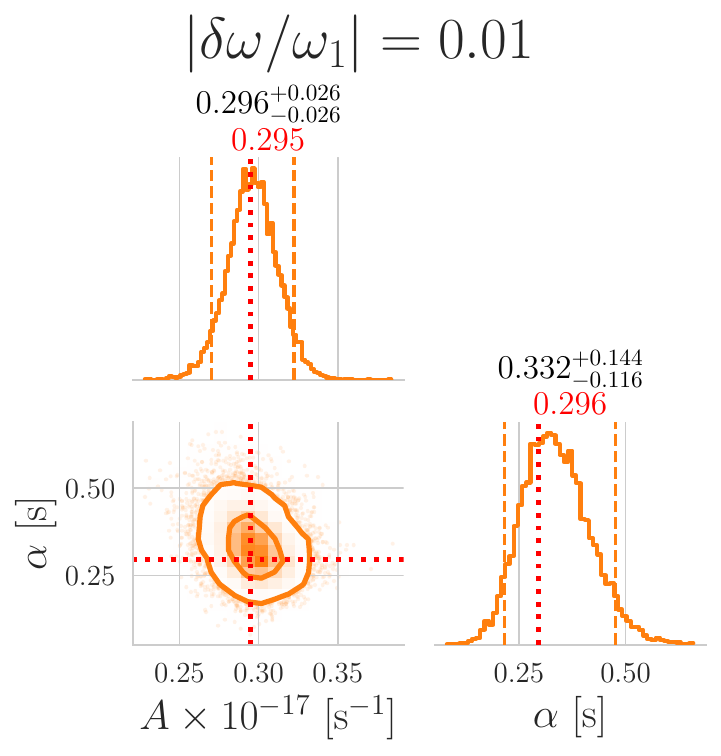}
      \end{minipage}
      \begin{minipage}[t]{0.28\textwidth}
        \centering
        \includegraphics[keepaspectratio, width=1\linewidth]{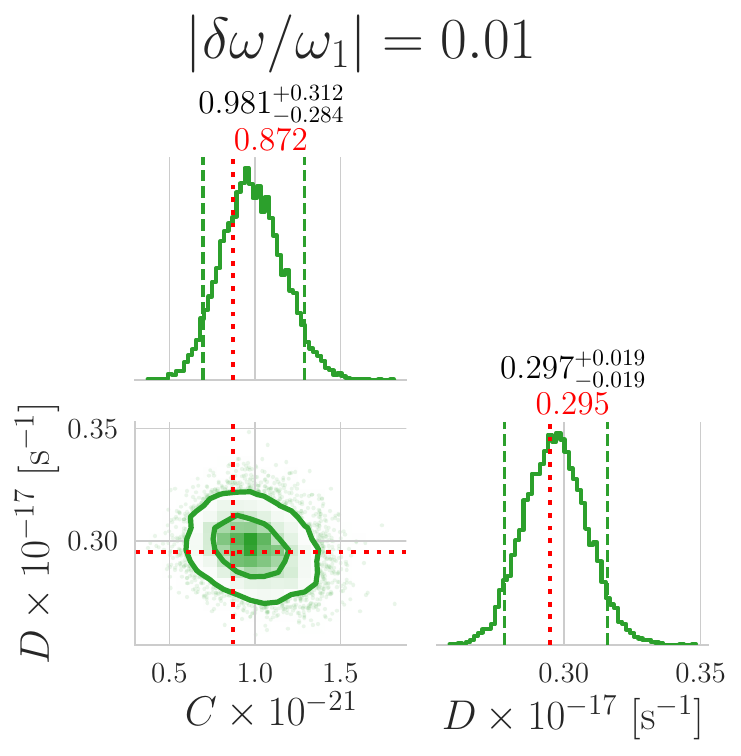}
      \end{minipage} \\
      \begin{minipage}[t]{0.28\textwidth}
        \centering
        \includegraphics[keepaspectratio, width=1\linewidth]{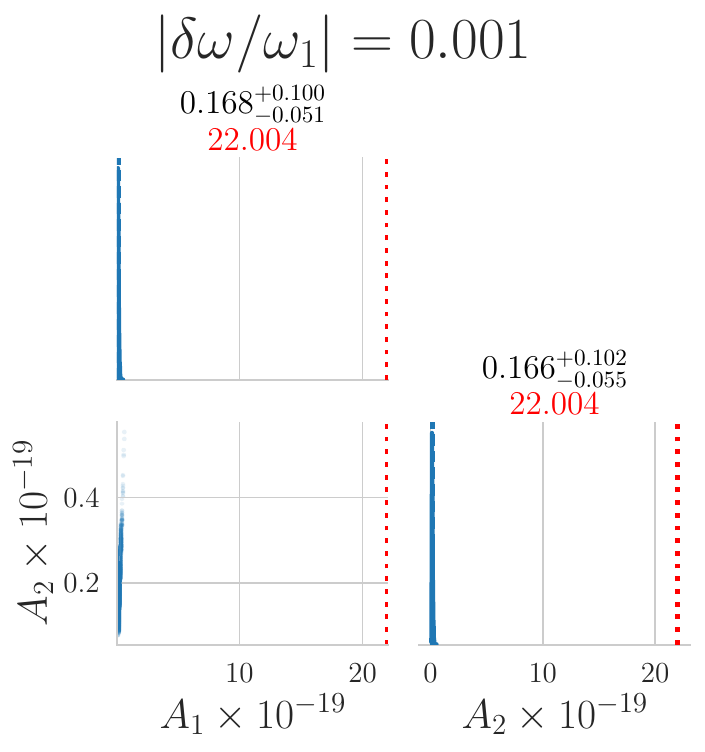}
      \end{minipage} &
      \begin{minipage}[t]{0.28\textwidth}
        \centering
        \includegraphics[keepaspectratio, width=1\linewidth]{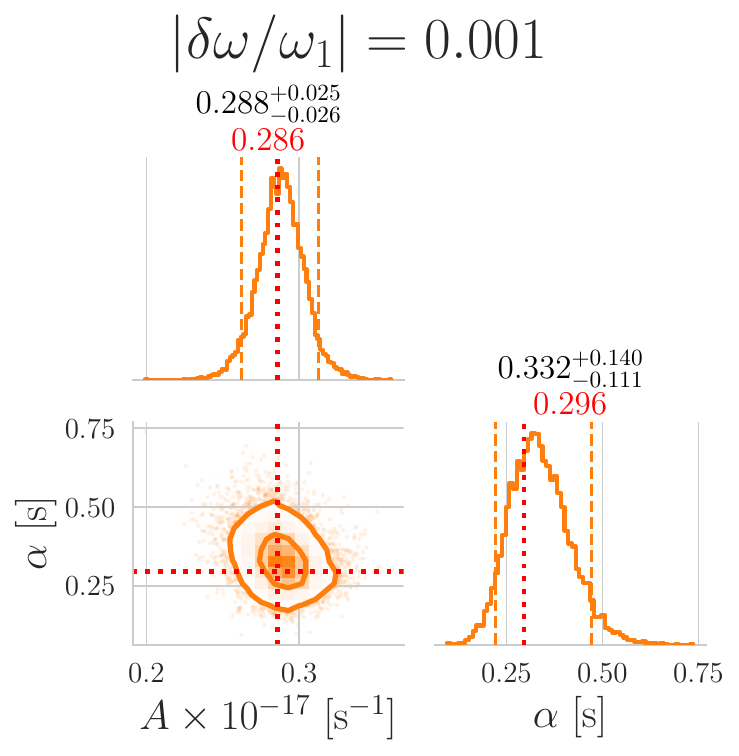}
      \end{minipage}
      \begin{minipage}[t]{0.28\textwidth}
        \centering
        \includegraphics[keepaspectratio, width=1\linewidth]{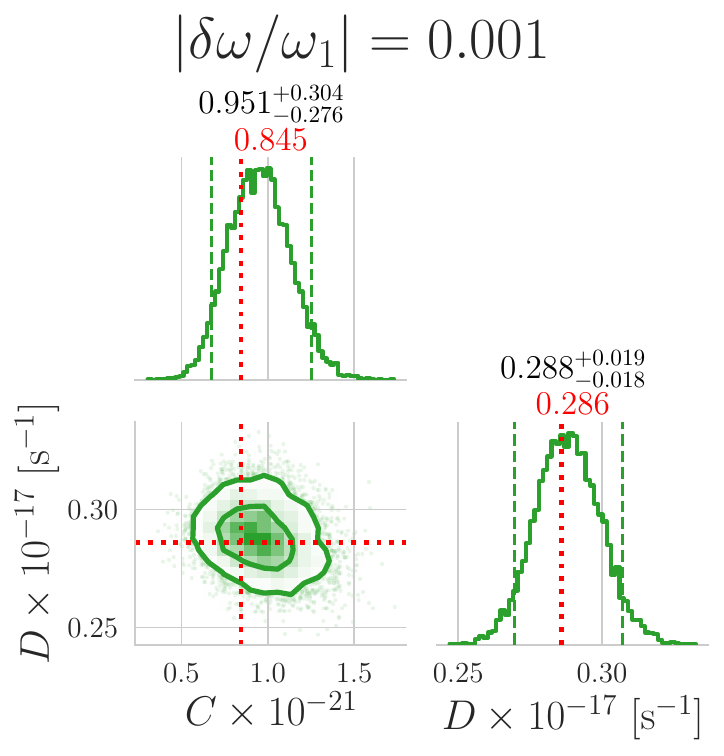}
      \end{minipage}
    \end{tabular}
    \caption{Similar to Fig.~\ref{fig:corner plot for amplitudes in frequency shifted injection}, but for the damping-time-shifted injections.
    }
    \label{fig:corner plot for amplitudes for damping time shifted injection}
\end{figure*}

Figures~\ref{fig:corner plot for amplitudes in frequency shifted injection} and \ref{fig:corner plot for amplitudes for damping time shifted injection} present the posterior distributions of the amplitude parameters for the frequency-shifted injection and damping-time-shifted injections, respectively. 
The 2DS model, shown in the left columns, exhibits systematic underestimation as $|\delta\omega/\omega_1|$ deceases, consistent with the trend summarized in Fig.~\ref{fig:fractional accuracy for amplitudes}. 
The 2AC model yields posteriors centered around the injected values, while the EP model provides a good approximation only for sufficiently small separations because of the limited validity of the double-pole approximation.

\bibliography{reference}

\end{document}